\documentclass{vldb}

\usepackage{mathtools}
\usepackage{graphicx}
\usepackage{url}
\usepackage[T1]{fontenc}
\usepackage{epstopdf}
\usepackage{multirow}
\usepackage{dblfloatfix}
\usepackage{wrapfig}

\usepackage{graphicx}
\usepackage{balance}  
\usepackage{subfigure}
\usepackage{enumitem}
\usepackage{color}

\usepackage{algorithm}
\usepackage{algorithmicx}
\usepackage[noend]{algpseudocode} 
\usepackage{url}
\usepackage{amsmath,amssymb,bm,tensor,varwidth}

\newcommand\norm[1]{|#1|}

\algnewcommand{\IIf}[1]{\State\algorithmicif\ #1\ \algorithmicthen}
\algnewcommand{\EIf}[1]{\State\algorithmicelse\  \algorithmicif\ #1\ \algorithmicthen}

\newtheorem{theorem}{Theorem}
\newtheorem{definition}{Definition}
\newtheorem{lemma}{Lemma}

\newtheorem{problem}{Problem}

\newcommand{\E}{\ensuremath{\mathrm{E}}}

\newcommand{\NI}{\textsf{NI}}
\newcommand{\RBA}{\textsf{GDB}}
\newcommand{\GDB}{\textsf{GDB}}
\newcommand{\EMD}{\textsf{EMD}}

\newcommand{\LP}{\textsf{LP}}

\newcommand{\flickr}{\textsf{Flickr}}
\newcommand{\twitter}{\textsf{Twitter}}

\newcommand{\squishlist}{
 \begin{list}{$\bullet$}
  {  \setlength{\itemsep}{0pt}
     \setlength{\parsep}{3pt}
     \setlength{\topsep}{3pt}
     \setlength{\partopsep}{0pt}
     \setlength{\leftmargin}{2em}
     \setlength{\labelwidth}{1.5em}
     \setlength{\labelsep}{0.5em}
} }
\newcommand{\squishlisttight}{
 \begin{list}{$\bullet$}
  { \setlength{\itemsep}{0pt}
    \setlength{\parsep}{0pt}
    \setlength{\topsep}{0pt}
    \setlength{\partopsep}{0pt}
    \setlength{\leftmargin}{2em}
    \setlength{\labelwidth}{1.5em}
    \setlength{\labelsep}{0.5em}
} }

\newcommand{\squishdesc}{
 \begin{list}{}
  {  \setlength{\itemsep}{0pt}
     \setlength{\parsep}{3pt}
     \setlength{\topsep}{3pt}
     \setlength{\partopsep}{0pt}
     \setlength{\leftmargin}{1em}
     \setlength{\labelwidth}{1.5em}
     \setlength{\labelsep}{0.5em}
} }

\newcommand{\squishend}{
  \end{list}
} 
\newenvironment{myitemize}
{
    \begin{list}{\labelitemi}{\leftmargin=1em}
        \setlength{\topsep}{0pt}
        \setlength{\parskip}{0pt}
        \setlength{\partopsep}{0pt}
        \setlength{\parsep}{0pt}
        \setlength{\itemsep}{0pt}
}
{
    \end{list}
}

\begin{document}

\title{Uncertain Graph Sparsification}

\numberofauthors{1} 

\author{
\begin{tabular}{lllll}
\hspace{-0.8cm}
Panos Parchas\textsuperscript{1} & Nikolaos Papailiou\textsuperscript{2} & Dimitris Papadias\textsuperscript{1}  & Francesco Bonchi\textsuperscript{3}
\end{tabular}
\\[3mm]
\begin{tabular}{cccc}
\hspace{-0.8cm}
    \affaddr{\textsuperscript{1}UST, Hong Kong}
    &\affaddr{\textsuperscript{2}NTUA, Greece}
    &\affaddr{\textsuperscript{3}Eurecat, Spain}\\
    \affaddr{\{pparchas, dimitris\}@cse.ust.hk}
    &\affaddr{npapa@cslab.ece.ntua.gr}
    &\affaddr{francesco.bonchi@isi.it}\\ 
\end{tabular}}

\maketitle

\vspace{-10in}
\begin{abstract}
Uncertain graphs are prevalent in several applications including communications systems, biological databases and social networks. The ever increasing size of the underlying data renders both graph storage and query processing extremely expensive. Sparsification has often been used to reduce the size of deterministic graphs by maintaining only the important edges. However, adaptation of deterministic sparsification methods fails in the uncertain setting. To overcome this problem, we introduce the first sparsification techniques aimed explicitly at uncertain graphs. The proposed methods reduce the number of edges and redistribute their probabilities in order to decrease the graph size, while preserving its underlying structure. The resulting graph can be used to efficiently and accurately approximate any query and mining tasks on the original graph. An extensive experimental evaluation with real and synthetic datasets illustrates the effectiveness of our techniques on several common graph tasks, including clustering coefficient, page rank, reliability and shortest path distance.
\end{abstract}

\section{Introduction} \label{sec:intro}
Uncertain graphs, where edges are associated with a probability of existence, have been used widely in numerous applications. For instance, in communication systems, each edge $(u,v)$ is often associated with a reliability value that represents the probability that the channel from $u$ to $v$ will not fail. In biological databases, uncertain edges between vertices representing proteins are due to error-prone laboratory measurements. In social networks, edge probabilities can model the influence between friends, or the likelihood that two users will become friends in the future. 

\begin{figure}[t]
    \begin{center}
        \centering
        \subfigure[uncertain graph $\mathcal{G}$]{
            \includegraphics[width=0.4\linewidth]{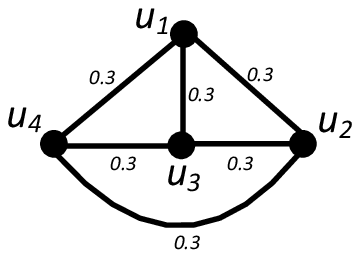}
        }
        \subfigure[sparsified graph $\mathcal{G'}$]{
            \includegraphics[width=0.4\linewidth]{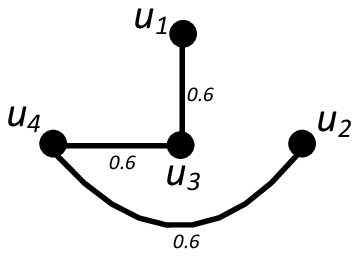}
        }
\vspace{-0.1cm}
        \caption{Uncertain graph sparsification example \label{fig:example1}}
            \end{center}
\vspace{-0.4cm}
\end{figure}

Several techniques have been proposed for diverse query processing and mining tasks on uncertain graphs (e.g. \cite{cliquesICDE15,bonchi2014core,hintsanen2008finding,PBGK10}), most of which assume \emph{possible-world} semantics. Specifically, let $\mathcal{G} = (V,E,p)$ be an uncertain (also called probabilistic) graph\footnote{We assume that $\mathcal{G}$ is unweighted, undirected and connected.}, where $p:E \rightarrow (0,1]$ assigns a probability to each edge. $\mathcal{G}$ is interpreted as a set $\{G = (V, E_G)\}_{E_G \subseteq E}$ of $2^{|E|}$ possible deterministic graphs, each defined on a subset of $E$. For example, since the uncertain graph of Figure \ref{fig:example1}(a) consists of 6 edges, there are $2^6$ possible worlds. 
Under this interpretation, exact processing requires query evaluation on all possible worlds and aggregation of the partial results. In general, the probability of a query predicate $Q$ is derived by the sum of probabilities of all possible worlds $G$ for which $Q(G) = true$:
\begin{equation} \label{equ:example}
Q(\mathcal{G}) = \sum_{\substack{G\sqsubseteq \mathcal{G}, \\ Q(G) = true}} \Pr(G)
\end{equation}
Applying Equation (\ref{equ:example}), the probability that the uncertain graph of Figure \ref{fig:example1}(a) contains a single connected component is $\Pr[\mathcal{G}$ is connected]=$0.219$. This  is obtained by generating the $2^6$ deterministic graphs, and adding the probabilities of the connected~ones.

Consequently, exact processing is prohibitive even for uncertain graphs of moderate size due to the exponential number of worlds. Thus, most techniques provide approximate results by applying Monte-Carlo (MC) sampling on a random subset of possible worlds. However, even MC may be very expensive for large uncertain graphs because generating a sample is time consuming as it involves sampling each edge. Moreover, due to the high entropy\footnote{ The entropy $H(\mathcal{G)}$ of an uncertain graph $\mathcal{G}$ is defined as the joint entropy of its edges $H(e)$ for all $e\in E$. Due to the edge independence, $H(\mathcal{G})=\sum_{e\in E}{H(e)}=\sum_{e\in E}(-p_e\log p_e-(1-p_e)\log (1-p_e))$.} of the uncertain graphs, there is significant variance among the possible worlds, which implies the need of numerous samples for accurate query estimation. This imposes huge overhead at query processing cost, because the query must be executed at every sample.

In order to tackle the high cost, we develop techniques for \textit{uncertain graph sparsification}. Specifically, given $\mathcal{G}$ and a parameter $\alpha \in (0,1)$, the proposed methods generate a sparsified probabilistic subgraph $\mathcal{G}' = (V,E',p')$, which contains a fraction of the edges, i.e.,
$E' : E' \subset E$, $|E'| = \alpha|E|$. 
$\mathcal{G}'$ preserves the structural properties of $\mathcal{G}$, has less entropy, and can be used to approximate the result of a wide range of queries on $\mathcal{G}$. Sparsification yields significant benefits in execution time because the cost of sampling is linear to the number of edges. Moreover, the required number of samples is proportional to the graph's entropy (~\cite{dang2015dimension,varianceReduction}), which is lower in the sparcified graph. Finally, similar to the case of deterministic graphs, sparsification reduces the storage cost, and facilitates visualization of complex networks.
 
Figure \ref{fig:example1}(b) illustrates $\mathcal{G'}$, a sparsified subgraph of $\mathcal{G}$ that contains half the original edges. Observe that the edges of $\mathcal{G'}$ have higher probabilities than those in $\mathcal{G}$, in order to compensate for the missing ones. Assume, for instance, a query that asks for the probability that $\mathcal{G}$ consists of a single connected component. Since $\Pr[\mathcal{G'}\mbox{ is connected]=0.216}$ and $\Pr[\mathcal{G}\mbox{ is connected]=0.219}$, $\mathcal{G'}$ can be used to effectively approximate the result of the query. Furthermore, the entropy decreases from 0.94 to 0.4. Since, $\mathcal{G'}$ has fewer edges, it is more efficient to sample and store. Additionally, since it has less entropy, it requires fewer samples for accurate estimation. Our goal is to generate sparsified graphs that can be used for a variety of queries and tasks. 

To the best of our knowledge, this is the first work on uncertain graph sparsification. On the contrary, sparsification has received considerable attention in the deterministic graph literature. In that context, most techniques aim at approximating all shortest path distances up to a multiplicative or additive factor, or preserving all cuts up to an arbitrarily small multiplicative error. As we demonstrate in our experimental evaluation, the adaptation of such methods to uncertain graphs yields poor results. On the other hand, our sparsification techniques achieve high accuracy and small variance for common graph tasks by capturing the \textit{expected} node degrees, or the \textit{expected} cut sizes up to a certain value. Summarizing, the contributions of the paper are:
\begin{myitemize}
\item We propose a novel framework of uncertain graph sparsification with entropy reduction.
\item We design algorithms that reduce the number of edges and tune the probability of the remaining ones to preserve crucial properties.
\item We experimentally demonstrate that the sparsified graphs are effective for a variety of common tasks including \textit{shortest path distance}, \textit{reliability}, \textit{page rank} and \textit{clustering coefficient}.
\end{myitemize}

The rest of the paper is organized as follows. Section 2 surveys the related work. Section 3 defines the problem, introduces our uncertain sparsification framework and presents baseline solutions motivated by the deterministic graph literature. Section 4 proposes sparsification algorithms that capture the expected vertex degrees. Section 5 analyzes rules to preserve cut sizes and modifies our algorithms for this case. Section 6 contains an extensive experimental evaluation on real and synthetic datasets, and Section 7 concludes the paper.
\vspace{3cm}
\section{Related Work} \label{sec:rel}
Existing sparsification methods focus exclusively on deterministic graphs.
Section \ref{subsec:rwdgs-sp}  and  \ref{subsec:rwdgs-cuts} present sparsification techniques that preserve the shortest path distances and  graph cuts respectively. Section \ref{subsec:rwdgs-general} discusses other related methods.

\subsection{Spanners}\label{subsec:rwdgs-sp}
Sparsification aimed at preserving the shortest path distances between all node pairs is based on the concept of $t$-$spanner$ \cite{peleg1989graph}. A $t$-$spanner$ of a weighted  deterministic graph $G = (V,E,w)$ is a subgraph $G'=(V, E',w), E' \subseteq E$ such that, for any pair of vertices $u,v \in V$, their distance in $G'$ is at most $t\in \mathbb{N}^+$ times their distance in $G$, i.e., $dist_{G'}(u,v)\leq t\cdot dist_G(u,v)$. The parameter $t$ is the \textit{stretch factor} of $G'$. Computing a  $t$-$spanner$ with the minimum number of edges has received considerable attention in the literature \cite{ruan2011distance}. Peleg and  Sch{\"a}ffer \cite{peleg1989graph} prove that 
a spanner with stretch factor $2t-1$ must have at least $\Omega(n^{1+1/t})$ edges.

Based on this lower bound,
a simple technique \cite{althofer1993sparse} to generate $(2t-1)$-$spanners$ processes the edges in increasing order of their weights. Initially, the spanner is $E' = \emptyset$. An edge $e=(u, v)$ is included to $E'$, if the distance between $u$ and $v$ in $E'$ exceeds $(2t-1) w_e$. The algorithm has  $O(|E| n^{1+1/t})$ time complexity due to the shortest path distance computations. Roditty and Zwick \cite{roditty2004dynamic} propose an improved $O(t n^{2+1/t})$ time method, which incrementally maintains a single source shortest path tree up to a given distance. Finally, Baswana et. all \cite{baswana2007simple} design a randomized algorithm to compute a spanner of $(2t-1)$-stretch and $O(t |E|^{1+1/t})$ size in $O(t|E|)$ expected time, which uses a novel clustering approach and avoids distance computations. An adaptation of this method provides a benchmark for our experimental evaluation.

\subsection{Cut-based sparsifiers}\label{subsec:rwdgs-cuts}

Let a deterministic, undirected, weighted graph $G=(V,E,w)$, where $|V|=n$. Given a vertex set $S \subseteq V$, the cut $E_G(S)$ is the set of edges with exactly one endpoint in $S$, i.e, $E_G(S)=\{(u,v) \in E | (u \in S, v \notin S)\}$. The size $C_G(S)$ of the cut is the sum of weights of the edges in $E_{G}(S)$, i.e., $C_G(S)=\sum_{e \in E_G(S)} w_e$. Cut-based sparsification preserves the size of \textit{all} cuts of the graph within a multiplicative error. Formally, given a dense weighted graph ($|E|=\Theta(n^2)$) and an approximation error $\epsilon\in (0,1)$, the output is $G'=(V,E',w')$, where $E'\subseteq E$ and  $|E'| = O(n\log n/\epsilon^2)$, such that for any set $S\subseteq V$, $C_{G{'}}(S)$ is within $\epsilon$ of the original cut size $C_G(S)$, i.e., $C_{G{'}}(S)\in (1\pm \epsilon)C_G(S)$ with high probability.

Algorithms for cut-based sparsification follow a two step approach. 
The first assigns a probability $p_e$ to each edge based on the topology of the graph. The second step samples each edge with probability $p_e$, and assigns weight $w'_e \propto \frac{1}{p_e}$ to the sampled edges. Intuitively, sparsification through this framework relies on the following observations: 
\begin{myitemize}
\item Edges in dense areas are not crucial for maintaining the graph connectivity. Thus, they have low sampling probability $p_e$ \cite{fung2011general}. 
\item Edges sampled with low $p_e$ are assigned large weights in order to compensate for their missing neighbouring edges.
\end{myitemize}
The existing methods assume integer weights and differ mainly in the first step, i.e., that of choosing $p_e$ for each edge $e=(u,v)$.  Bencz\'{u}r and Karger \cite{benczur1996approximating} assign probabilities inversely proportional to the $k$-strong connectivity\footnote{A maximal $k$-strong connected component is a maximal induced subgraph of $V$ that remains connected after removing up to $k$ edges. The $k$-strong connectivity of an edge $(u,v)$ is the maximum $k$ such that $u$ and $v$ belong to the same $k$-strong connected component.} of $u,v$.  Fung et al. \cite{fung2011general} simplify analysis by utilizing sampling probabilities inversely proportional to the size of the minimum cut separating $u$ and $v$. 
Nagamochi and Ibaraki \cite{nagamochi1992linear} estimate the edge connectivities using the NI index $\lambda_e$ of an edge $e$. The NI index is generated by iteratively constructing spanning forests of the initial graph, while reducing the weights of the selected edges. In essence, the NI index is the last spanning forest that contains $e$, given that an edge with weight $w_e$ needs to participate in $w_e$ contiguous forests. To ensure that the sparse graph has $O(n\log n/\epsilon^2)$ edges in expectation, the sampling probability is set to $p_e=\frac{\rho}{\lambda_e}$, where $\rho=O(\log n/\epsilon^2)$. To see this, let $\mathbb{E}(|E'|)$ denote the expected number of edges of $G'$. According to \cite{benczur2015randomized}, $\sum_e{1/\lambda_e}=O(n\log n)$; thus, $\mathbb{E}(|E'|)=\sum_e{p_e}=\rho\sum_e {1/\lambda_e}=O(n\log^2 n/\epsilon^2)$. A more refined analysis \cite{fung2011general} reduces this to $O(n\log n/\epsilon^2)$. Section \ref{seq:meth-0} modifies \cite{nagamochi1992linear} as a benchmark for our experiments.

Spielman and Srivastava \cite{spielman2008graph} generate the electrical equivalent of the graph by assuming resistors of 1$\Omega$ at each link. The sampling probability of edge $(u,v)$ is proportional to the amount of current that flows through $e$ when a unit voltage difference is applied to $u,v$. This approach also preserves every eigenvalue of the original graph, leading to the stronger notion of \textit{spectral sparsification}\cite{spielman2008graph}. 
\cite{Kapralov2013Spectral, Koutis2014Simple} employ spanners in the context of spectral sparsification. \cite{Kapralov2013Spectral} utilizes spanners to approximate the "robust connectivity" $q_\kappa(u,v)$, i.e., the number of paths between vertices $u$ and $v$ with length at most $\kappa$. It is proven that $q_\kappa(u,v)$ serves as a good upper bound for the effective resistance of \cite{spielman2008graph}. On the other hand, \cite{Koutis2014Simple} extracts a collection of edge disjoint spanners and then samples the remaining edges with fixed probability, yielding a parallel algorithm that spectrally preserves the graph. Furthermore, spectral sparsification algorithms have been proposed for the streaming (\cite{Ahn2012Graph,Ahn2012Spectral,Kapralov2017Single}) and dynamic \cite{Abraham2016OnFully} graph settings. For a more detailed review of cut-based  sparsification literature on deterministic graphs, we refer the reader to \cite{Batson2013Spectral}.


\subsection{Other related techniques}\label{subsec:rwdgs-general}
Other than the theoretical papers presented above, there has been little work on implementing sparsification algorithms for deterministic graphs. Linder et al.\cite{Linder2015Structure} propose some heuristic methods aiming to preserve the structure of social networks. Their main method \textit{Local Degree} keeps only the edges to \textit{hubs}, i.e., vertices with high degree, claiming that they are crucial for the topology of complex social networks. However, the approach applies only on unweighted graphs and does not perform weight redistribution among the remaining edges; thus, it cannot be adapted to uncertain graphs.

Some deterministic graph sparsification techniques focus on approximating the result of particular queries, as opposed to structural graph properties. For instance, \cite{mathioudakis2011sparsification, bonchi2013activity} sparsify a directed graph, while preserving the influence logs among its vertices. Satuluri et al. \cite{Satuluri:2011:LGS:1989323.1989399} apply local sparsification in order to capture communities and facilitate clustering.  These approaches are specific to the query in question, whereas we aim at generating sparsified graphs applicable to multiple, possibly diverse, query types. \textit{Subgraph sampling} generates a subgraph of an input deterministic graph that contains a small fraction of the nodes (and edges), but has similar structural properties \cite{Leskovec2006, HublerICDM08}. This is different from sparsified graphs, which maintain the input set of nodes. In our experiments, we apply \cite{Leskovec2006} to reduce the size of real graphs for expensive queries that cannot terminate in the original graphs.

\cite{parchas2014pursuit,parchas2015uncertain} propose algorithms for generating deterministic representative instances that approximate the expected node degrees of the uncertain graph. Queries can then be processed by applying conventional algorithms on these instances. Since the representatives have fewer edges than the original uncertain graph, this could be viewed as a special case of sparsification. However, a representative (i.e., deterministic graph) cannot be used to answer queries whose output is uncertain, e.g., return the probability that the graph consists of a single connected component, or the probability that two vertices are reachable from each other. On the other hand, our techniques generate uncertain graphs that can be used for the same query and mining tasks as the original graph. Moreover, the methods of \cite{parchas2014pursuit,parchas2015uncertain} do not provide control over the number of edges in the representative graphs. Intuitively, while \cite{parchas2014pursuit,parchas2015uncertain} aim at \textit{eliminating} uncertainty by extracting a representative instance (i.e., zero entropy), in this work we aim at \textit{decreasing} the uncertainty of the input graph (i.e., reducing its entropy).

%
\section{Problem Definition and Framework} \label{sec:probdef}

Let $\mathcal{G}=(V,E,p)$ be a probabilistic undirected graph, where $p: E \rightarrow (0,1]$ is a function that assigns a probability $p(u,v)$ to each edge $(u,v) \in E$.
Given a \textit{sparsification ratio} $\alpha \in (0,1)$, we extract from $\mathcal{G}$ a sparsified subgraph $\mathcal{G}' = (V,E',p')$ such that $E' \subset E$ and $|E'| = \alpha|E|$. $\mathcal{G}'$ should preserve the structural properties of $\mathcal{G}$, so that it can be used to accurately approximate the result of diverse queries on $\mathcal{G}$. Moreover, $\mathcal{G'}$ should reduce the entropy of $\mathcal{G}$ in order to decrease the variance of the queries.  In addition to diminishing the storage overhead, sparsification yields significant benefits in terms of query processing because the cost of sampling is proportional to the number of edges\footnote{Sampling techniques have complexity $O(E)$ \cite{li2014ICDE}.}. Moreover, through entropy reduction, the resulting graph is less uncertain, thus it requires fewer samples for accurate query estimation.

\subsection{Uncertain sparsification}

As stated in Section 2, a prevalent goal of deterministic graph sparsification is preservation of the cut sizes. The notion of a cut can be extended naturally to uncertain graphs. In this case, due to the linearity of expectation, the \textit{expected} size of a cut is the sum of the probabilities of the edges involved in the cut.
\begin{definition}[Expected cut size]
Given an uncertain\\ graph $\mathcal{G} = (V,E,p)$ and a subset $S\subseteq V$, the expected cut size of $S$ in $\mathcal{G}$ is the summation of the probabilities of the edges with exactly one endpoint in $S$:
\vspace{-0.1cm}
$$
C_\mathcal{G}(S)=\sum\limits_{\substack{e=(u,v) \in E \\ u \in S, v \in V \setminus S}}p_e
$$
\end{definition} 
\vspace{-0.2cm}
We define the \textit{absolute discrepancy} $\delta_A(S)$
of a vertex set $S$ in a sparsified graph $\mathcal{G'}$ as the difference of $S$'s expected cut size in $\mathcal{G'}$ to its expected cut size in $\mathcal{G}$, i.e., 
\vspace {-0.05in}
\begin{equation*}
\delta_A(S)= C_\mathcal{G}(S)-C_\mathcal{G'}(S) 
\end{equation*}
Accordingly, the \textit{relative discrepancy} $\delta_R(S)$ is the absolute discrepancy of $S$ divided by the original cut size:
\begin{equation*}
\delta_R(S)= \frac{C_\mathcal{G}(S)-C_\mathcal{G'}(S)}{C_\mathcal{G}(S)}
\end{equation*}
To simplify notation, we collectively refer to $\delta_A$ and $\delta_R$ as $\delta$, and we only differentiate when required. In addition, we use the term cut to also refer to the expected cut size. Motivated by the work in deterministic graph sparsification, we aim at cut-preserving sparsified graphs, or, using our notation, at minimizing discrepancy $\delta$. The exponential number of cuts renders their exhaustive enumeration intractable. To overcome this, we target cuts of sets $S$ with specific cardinality $k$.

Formally, given an integer $k$, we define the $k$-discrepancy $\Delta_k$ of a graph $\mathcal{G}'$ as the sum of the absolute values of the discrepancies for all sets with cardinality $k$:
\vspace{-0.2cm}
\begin{equation*}\label{costf}
\Delta_k(\mathcal{G'})=\sum_{\substack{S \subseteq V},|S|=k} |\delta(S)| 
\end{equation*}
where $\delta(S)$ is the absolute or relative discrepancy of $S$. The absolute discrepancy emphasizes vertices with high degree because they are more likely to yield large absolute errors. On the other hand, the relative discrepancy targets all node degrees equally, by considering the relative error. We aim at minimizing the sum of $\Delta_i$ for $1 \leq i \leq k$, or equivalently at preserving the size of all cuts up to $k$. Accordingly, the problem we tackle in this work is:

\begin{problem}\label{Prob1}
Given an uncertain graph $\mathcal{G} = (V,E,p)$, and a sparsification ratio $\alpha\in (0,1)$, find an uncertain graph $\mathcal{G}^*= (V,E^*,p^*)$, with $|E^*|=\alpha|E|$ that minimizes the sum of discrepancies $\sum_{i=1}^{k}{\Delta_i(\mathcal{G^*})}$ up to a given $k\geq 1$ and the entropy $H(\mathcal{G^*})$.
\end{problem}
 
For the special case of $k=1$, minimizing $\Delta_1$ is equivalent to preserving the expected degrees for all vertices, which has been shown to be effective when generating deterministic representatives of uncertain graphs \cite{parchas2014pursuit}. 

\subsection{Benchmark solutions}
\label{meth:cuts}
Spanners and cut based sparsifiers stem from theoretical papers, and to the best of our knowledge, they have not been implemented or evaluated in practice. Furthermore, our uncertain graph sparsification setting differs from that of deterministic sparsification. All spanners focus on selecting a subset of edges without changing their weights. On the other hand, we modify the probabilities of edges in the sparsified graph, in order to compensate for the eliminated edges. Moreover, probabilities of the uncertain graphs, unlike weights of deterministic graphs are bounded by 1, inhibiting the direct application of cut-based sparsification methods, all of which assume unbounded weights. Lastly, uncertain sparsification explicitly targets to reduce the entropy of the input graph in order to minimize the variance of query estimators. Deterministic sparsifiers do not consider entropy reduction. Nevertheless, 
in the following, we extend two state-of-the-art methods, one based on \textit{cut sparsifiers} and the other on \textit{spanners}, to uncertain graphs. The resulting algorithms are used as benchmarks in our experiments.  

As a representative of cut sparsifiers, we adopt the NI method\footnote{Any method of Section \ref{subsec:rwdgs-cuts} can be applied similarly.} of \cite{nagamochi1992linear} by transforming the uncertain graph $\mathcal{G}$ to a weighted deterministic $G_w$. Recall that NI requires integer weights and an approximation parameter $\epsilon \in (0,1)$ to produce a sparsified graph $G'_w$ with $O(n\log n/\epsilon^2)$ edges on expectation. Intuitively, the probabilities of $\mathcal{G}$ are directly analogous to the weights of $G_w$ in terms of (expected) cut size. 
To maintain this analogy while ensuring $w_e \in \mathbb{N}^+$, our transformation divides each probability of $\mathcal{G}$ by the smallest value $p_{min}$, and rounds the result to the closest integer, i.e., $w_e=\lfloor p_e/p_{min}\rceil$. 
To relate $\epsilon$ to our sparsification ratio $\alpha$, we set $\epsilon=\sqrt{n\log n/\alpha|E|}$. Next, we use NI as a black box to sparsify $G_w$ into $G'_w$.
However, since the number of edges of $G'_w$ in \cite{nagamochi1992linear} is given on expectation with asymptotic notation, it is not guaranteed to equal $\alpha|E|$. If the resulting graph has more (fewer) edges than $\alpha|E|$, we iteratively execute NI after increasing (decreasing) $\epsilon$ by a small factor $\theta$, until the first (last) graph for which $|E'|\leq \alpha|E|$. The remaining $\alpha|E|-|E'|$ edges are randomly sampled from $E \setminus E'$ using the initial probabilities $p$. Intuitively, this calibration process approximates the minimum $\epsilon$, which ensures $|E'|\leq \alpha|E|$. 
Finally, for each edge $e$, we convert $w'_e$ into $p'_e$ through the inverse transformation cupped by 1, i.e, $p'_e=\min\{w'_e\cdot p_{min},1\}$. 

In order to apply spanners, we generate the weights of $G_w$ using the formula $w_e=-\log(p_e)$ \cite{PBGK10}. This transformation takes advantage of the logarithmic summation properties to preserve the \textit{most probable paths} of $\mathcal{G}$. For the tuning of the stretch factor, we solve the equation $\alpha|E| =  tn^{1+1/t}$ with respect to $t$ in order to find the minimum spanner with the required number of edges $\alpha|E|$. Using the computed value of $t$, we run the spanner algorithm of \cite{baswana2007simple} on top of $G_w$. As in the case of cuts, some calibration steps may be required to approximate the minimum $t_{min}$ that ensures $|E'|\leq \alpha|E|$. This time however, at each iteration, $t_{min}$ can only change by 1 because it is integer. After computing $G'_w=(V,E',w)$, we add its edges in the sparse probabilistic graph $\mathcal{G'}$ using the initial probabilities $p'=p$, since \cite{baswana2007simple} retains the original edge weights. The appendix contains details and pseudocodes of the benchmark adaptations.

\subsection{Framework}\label{seq:meth-0}
The proposed framework starts with an initialization step that generates a connected unweighted backbone graph $G_b$. Then two different techniques operate on $G_b$ in order to produce the sparsified graph. \textit{Gradient Descent Backbone} (\GDB) assigns probabilities to the edges of $G_b$ without altering its structure, i.e., the resulting graph has the same edges as $G_b$. On the other hand, \textit{Expectation Maximization Degree} (\EMD) updates both the structure of $G_b$ and the edge probabilities. In the rest of this section we focus on the initialization step.


A simple approach could sample the edges of $\mathcal{G}$ in random order according to their probabilities, until it obtains $\alpha|E|$ edges\footnote{A similar approach has been applied in \cite{Linder2015Structure} for sparsification of deterministic graphs.}. However, this would not ensure the connectivity of $G_b$, especially for small $\alpha$\footnote{We assume $\alpha\geq\frac{|V|-1}{|E|}$, otherwise the sparsified graph cannot preserve the connectivity of the original.}. Disconnected graphs can introduce large errors on various queries such as shortest distances between vertices of different connected components. Moreover, individual vertices may become entirely disconnected, which would lead to high total discrepancy. 

To overcome this problem, we generate connected backbone graphs using the following method, which is inspired by the related work in deterministic sparsification \cite{nagamochi1992linear}\footnote{Other deterministic sparsification methods such as $t$-bundle \cite{Koutis2014Simple} or \textit{Local Degree} \cite{Linder2015Structure} could also be used.}. We first compute a maximum spanning tree of $\mathcal{G}$, where the probabilities act as weights. Then, we remove the tree edges from $\mathcal{G}$ and insert them to $G_b$. This process is repeated until $G_b$ contains $\alpha'|E|$ edges, where $\alpha'<\alpha$. Note that after the first application of the maximum spanning tree, $\mathcal{G}$ may become disconnected; thus, subsequent applications may return spanning forests instead of trees. Finally, the remaining $(\alpha-\alpha')|E|$ edges of $G_b$ are generated by random sampling based on the edge probabilities. Algorithm \ref{algo:BGG} illustrates backbone graph generation.
\begin{algorithm}[ht]
  \caption{Backbone Graph Initialization (\textsf{BGI})}
  \label{algo:BGG}
  \begin{algorithmic}[1]
    \Require uncertain graph $\mathcal{G}=(V,E,p)$, sparsification ratio $\alpha$, spanning ratio $\alpha'$
    \Ensure backbone graph $G_b=(V,E_b)$
    \State $E_b \gets$ maximum spanning tree of $E$
	\State $E\gets E\setminus E_b$
     \label{line_BGG:1}
	\While {$|E_b|<\alpha'|E|$}
		\State $F\gets$ maximum spanning forest of $E$
		\State $E_b\gets E_b\cup F$
		\State $E\gets E\setminus F$
	\EndWhile
	
	\While {$|E_b|<\alpha|E|$}
		\State sample edge $e\in E$ with probability $p_e$
		\If {$e$ is selected}
			\State $E_b\gets E_b\cup \{e\}$
			\State $E\gets E\setminus \{e\}$
		\EndIf
	\EndWhile
  \end{algorithmic}
\end{algorithm}  
 
Parameter $\alpha'$ tunes the number of edges obtained through spanning forests. Intuitively, generating all $\alpha|E|$ edges using spanning forests is not desirable, because all vertices would be treated equally, independently of their degrees. In our experiments, we set the value of $\alpha'$ so that it is the minimum of $0.5\alpha$ and the number of edges in the first six maximum spanning forests. 

Given the initial graph $G_b$, Section \ref{sec:deg} proposes algorithms that aim at minimizing the degree discrepancy $\Delta_1$, while Section \ref{sec:cuts} focuses on preserving the expected cuts i.e., $\Delta_k$ for $k>1$. In both cases probability values that would incur high entropy are avoided. 

\section{Preserving Expected Degrees}\label{sec:deg}
We first focus on Problem 1 for $k=1$, and describe methods to generate a sparsified graph $\mathcal{G'}$ that preserves the expected degrees of all vertices in $\mathcal{G}$. Section \ref{seq:meth-1} describes probability assignment that minimizes $\Delta_1$ using linear programming (\LP). Due to the inefficiency of \LP, on large graphs, Sections \ref{seq:meth-2} and \ref{seq:meth-3} propose \RBA\ and \EMD, which assign probabilities inspired by gradient descent and expectation maximization, respectively. 

 
\subsection{Optimal probability assignement for minimizing $\Delta_1$} \label{seq:meth-1}


Given the backbone graph $G_b$, we wish to compute the edge probabilities that minimize the discrepancy for $k=1$. Let $\mathbf{d}$ be a vector of size $|V|$ that contains the expected degrees of the original graph $\mathcal{G}$. We represent $G_b$ by an incidence matrix $\mathbf{A_b}$ of size $|V|\times |E_b|$. Using this notation, an equivalent formulation for minimizing the sum of absolute discrepancies $\delta^A$ is
\vspace {-0.08in}
\begin{align}\label{equ:LP_def}
\underset{\mathbf{p'}}{\text{min.}}& \quad \norm{\mathbf{d}-\mathbf{A_b}\mathbf{p'}} \\
s.t. & \quad \mathbf{p'}\in (0,1]^{|E_b|} \nonumber
\vspace {-0.2in}
\end{align}

\begin{lemma} \label{lemma:LP1}
For an incidence matrix $\mathbf{A_b}$, there is a probability assignment $\mathbf{p^*}$ that minimizes $\Delta_1$ for which the expected degree $d^*_u\leq d_u,$ $\forall u\in V$.
\end{lemma}
\vspace {-0.1in}
\textit{Proof}: Consider a probability assignment $\mathbf{p^*}$ that minimizes $\norm{\mathbf{d}-\mathbf{A_b}\mathbf{p^*}}$ and let $\mathbf{d^*}=\mathbf{A_b}\mathbf{p^*}$ be the new vector of expected degrees. For the sake of contradiction, assume also that $\mathbf{d^*}$ contains \textit{illegal} vertices, i.e., vertices with $d^*_u>d_u$. Each illegal vertex $u$ is adjacent to at least one \textit{legal} vertex $v$, with $d^*_v\leq d_v$, otherwise the assignment is not optimal (setting $p^*(u,v)=p(u,v)-\epsilon$, for a small $\epsilon>0$ yields a better result) . Let $\theta=d^*_u-d_u>0$. We prove that, if we subtract $\theta$ from $p(u,v)$, then the resulting probability assignment $\mathbf{p{'}}$:
\vspace {-0.1in}
\[
 p{'}_e =
  \begin{cases}
   p_e- \theta, & \text{if } e =(u,v),\\
   p_e,      & \mbox{otherwise}
  \end{cases}
\]
is also optimal. Let $\mathbf{d{'}}=\mathbf{A_bp{'}}$ be the corresponding vector of expected degrees.
Then,
\vspace {-0.03in}
\begin{equation}\label{equ:d_i}
 d{'}_i =
  \begin{cases}
   d^*_i-\theta=d_i, & \text{if } i = u, \\
   d^*_i-\theta, & \text{if } i = v,  \\
   d^*_i,  & \mbox{otherwise} \\
  \end{cases}
\end{equation}
Since $\mathbf{p^*}$ is optimal, $\sum_{i\in V}|d_i-d^*_i|$ is minimum. Then,
\vspace {-0.03in}
\begin{align}\label{equ:LP_o}
&\sum_{i\in V}\norm{d_i-d^*_i}=\sum_{\mathclap{i\in V\setminus\{u,v\}}}\norm{d_i-d^*_i}+\norm{d_u-d^*_u}+\norm{d_v-d^*_v}\nonumber \\
&=\sum_{\mathclap{i\in V\setminus\{u,v\}}}\norm{d_i-d{'}_i}+\norm{d_u-d_u-\theta}+\norm{d_v-d{'}_v-\theta}
\end{align}
Given that $v$ is \textit{legal}, $d_v-d^*_v\geq 0\xRightarrow{(\ref{equ:d_i})} d_v-d{'}_v-\theta\geq 0$.
Thus, Equation (\ref{equ:LP_o}) becomes:
\begin{align*}
\sum_{i\in V}\norm{d_i-d^*_i}&=\sum_{\mathclap{i\in V\setminus\{u,v\}}}\norm{d_i-d{'}_i}+\theta+ \norm{d_v-d{'}_v}-\theta=\sum_{i\in V}\norm{d_i-d{'}_i}
\end{align*}
Since $\sum_{i\in V}\norm{d_i-d^*_i}$ is minimum, 
$\sum_{i\in V}\norm{d_i-d{'}_i}$ is also minimum. Thus, $\mathbf{p{'}}$ is optimal.
By applying the above argument to all \textit{illegal} vertices we construct an optimal instance that contains only legal vertices. $\square$

Through Lemma \ref{lemma:LP1} we prove the following theorem.
\begin{theorem} \label{theorem:LP}
Given a backbone incidence matrix $\mathbf{A_b}$ and an expected degree vector $\mathbf{d}$, the optimal probability distribution $\mathbf{p^*}$ for degree discrepancy $\delta_A$, is the solution of the following LP:
\end{theorem}
\vspace{-3mm} 
\begin{align}\label{equ:th1}
\underset{\mathbf{p'}}{\text{max.}}& \quad \norm{\mathbf{p'}} \nonumber\\
s.t. & \quad \mathbf{A_b}\mathbf{p'}\leq \mathbf{d} \\
& \quad \mathbf{p'}\in (0,1]^{|E|} \nonumber
\vspace{-1mm}
\end{align}
\textit{Proof}:
According to Lemma \ref{lemma:LP1}, there exists an optimal assignment $\mathbf{p^*}$ for which $\mathbf{A_b}\mathbf{p^*}\leq \mathbf{d}$. We only need to prove that the objective function is equivalent to Equation (\ref{equ:LP_def}). Let $\mathbf{A_b}=[\mathbf{a}_{u_1}\quad \mathbf{a}_{u_2}\quad \dots \quad \mathbf{a}_{u_n}]^T$, i.e., the row vectors of matrix $\mathbf{A_b}$. Then
\begin{align*}
&\min \norm{\mathbf{d}-\mathbf{A_bp'}}= \min \sum_{u\in V}\big|d_u-\mathbf{a}_u \mathbf{p'}\big|\stackrel{(\ref{equ:th1})}{=}\min \sum_{u\in V}\big(d_u-\mathbf{a}_u \mathbf{p'}\big) \\
&= \max {\sum_{u\in V}{\mathbf{a}_u \mathbf{p'}}}=\max {\sum_{u\in V}{p_u}}= \max{\norm{\mathbf{p'}}} \quad _{\square}
\end{align*}
Theorem \ref{theorem:LP} states that the probability assignment step can be performed optimally by any linear programming solver, e.g. simplex. However, the running time of such solvers is prohibitive for large graphs, which is also confirmed in our experimental evaluation. Moreover, \LP\ does not explicitly reduce entropy. The following method closely approximates the optimal probability assignment at a small fraction of the time, while also decreasing the entropy compared to $\mathcal{G}$.

\subsection{Gradient descent backbone} \label{seq:meth-2}

Given the backbone graph $G_b=(V,E_b)$, \textit{Gradient Descent Backbone} (\GDB) initially generates a seed uncertain graph $\mathcal{\hat{\mathcal{G}}}=(V,E_b,\hat{p})$, $\hat{p}=p$, and proceeds in iterations. Let $\mathbf{\hat{p}}$ and $\mathbf{p'}$ be the probabilities of the previous and the current iteration, respectively. At each iteration, \GDB\ optimizes the probability $p'_e$ of each edge $e=(u_0,v_0)$, considering the remaining probabilities fixed. 
To this end, we need to calculate the derivative of the objective function $\sum_{u \in V} |\delta(u)|$. Since $|\delta(u)|$ is not differentiable, we utilize the squared discrepancy $\delta^2(u)$. Accordingly, the objective function of \GDB\ is $D_1=\sum_{u \in V} \delta^2(u)$. Its derivative with respect to $p'_e$ is:
\begin{equation}\label{equ:derivative}
\frac{\partial D_1'}{\partial p'_e}=\sum_{u\in V}{\frac{\partial \delta'^2(u)}{\partial p'_e}}=-2\cdot \delta'(u_0)- 2\cdot \delta'(v_0)	 
\end{equation}

Therefore, changing the probability of edge $e$ from $\hat{p}_e$ to $p'_e$ the discrepancy $\delta'(u)$ becomes:  
\begin{equation}\label{equ:delta_4}
\delta'(u)= \frac{\hat{\delta}_A(u) + (\hat{p}_e-p'_e)}{\pi(u)}
\end{equation}
where:
\vspace{-0.6cm}
\begin{equation*}\label{equ:poll}
\pi(u) = \left\{
  \begin{array}{lr}
   1 & \text{if } \delta(u) = \delta_{A}(u) \\
    C_{\mathcal{G}}(u)    & \text{if } \delta(u) = \delta_{R}(u) 
  \end{array}
  \right.
\end{equation*}
and $\delta_A$ and $\delta_R$ correspond to absolute and relative discrepancy, respectively. Substituting Equation (\ref{equ:delta_4}) to Equation (\ref{equ:derivative}), the probability that sets the first derivative to zero, is:  

\begin{equation}\label{equ:avgrule}
p'_e=\hat{p}_e+stp \mbox{, where } stp=\frac{\pi(v_0)\hat{\delta}_A(u_0)+\pi(u_0)\hat{\delta}_A(v_0)}{\pi(u_0)+\pi(v_0)}
\end{equation}
Equation (\ref{equ:avgrule}) raises two concerns: First,
probability $p'_{e}$ may fall outside the range [0,1]. 
In this case, $D_1$ is monotonic in [0,1] because it is convex (i.e.,$\frac{\partial^2 D_1}{\partial p^2_e}>0$). 
Second, the probability increase \textit{stp} may result to higher entropy for $e$, which is not desirable. \GDB\ overcomes these concerns by assigning probabilities using the following rule:
\begin{equation}\label{equ:rule}
	p'_e =  \tensor*[_0]{\Bigg\lfloor}{} \hat{p}_e+h\frac{\pi(v_0)\hat{\delta}_A(u_0)+\pi(u_0)\hat{\delta}_A(v_0)}{\pi(u_0)+\pi(v_0)}\Bigg\rceil^1
\end{equation}
where $\tensor*[_0]{\lfloor}{} x \rceil^1 = \max\{0,\min\{x,1\}\}$ and $h\in [0,1]$. 

In essence, \GDB\ performs gradient descent which is guaranteed to reach a local minimum of the objective function \cite{gradient}. Parameter $h$ relates the step size of gradient decent to the entropy. Intuitively, if the optimal assignment of Equation (\ref{equ:avgrule}), results in entropy increase, \GDB\ adds only a fraction $h$ of \textit{stp} to $p_e$, attenuating the negative side effect. This is a common practice in gradient decent techniques: instead of moving directly to the goal, move in smaller steps towards the correct direction \cite{gradient}. In addition to limiting entropy increase, this allows other neighbouring edges to update their probabilities, decoupling the local minimum from the edge ordering. 
   
Algorithm \ref{algo:RBA} contains the pseudocode of \GDB. Lines \ref{line_RBA:1}-\ref{line_RBA:2} initialize $\mathcal{G'}$ with all edges of the backbone graph $G_b$ using their corresponding probabilities in $\mathcal{G}$. Then, the algorithm iteratively examines every edge of $E'$ and decides its probability: if the optimal assignment of Equation (\ref{equ:avgrule}) leads to entropy increase, then $stp$ is capped by parameter $h$ (line \ref{line_RBA:4}). Otherwise, both the discrepancy and the entropy are reduced by the optimal assignment, and no limit is applied. The algorithm terminates when the improvement of the objective function is below a threshold $\tau$. 
\begin{algorithm}[!h]
  \caption{Gradient Descent Backbone (\GDB)}
  \label{algo:RBA}
  \begin{algorithmic}[1]
    \Require uncertain graph $\mathcal{G}=(V,E,p)$, backbone graph $G_b=(V,E_b)$, entropy parameter $h$
    \Ensure sparse uncertain graph $\mathcal{G'}=(V,E',p')$
    \State $E' \gets \emptyset$ \label{line_RBA:1}
	\ForAll {edge $e=(u,v)\in E_b$}	
		\State $ E' \gets E' \cup \{e\}$; $p'_e \gets p_e$
	\EndFor	\label{line_RBA:2}
	\Repeat	\label{line_RBA:3}
	\State $\hat{D}_{1} = D_{1}(\mathcal{G'})$
    	\ForAll {edge $e\in E'$}
    		\State \textit{stp}$\leftarrow \frac{\pi(v_0)\hat{\delta}_A(u_0)+\pi(u_0)\hat{\delta}_A(v_0)}{\pi(u_0)+\pi(v_0)};$ $p_e' \leftarrow p_e+stp$ 
    		\IIf {$p_e'<0$} $p_e'\leftarrow 0$ 
    		\EIf {$p_e'>1$} $p_e'\leftarrow 1$
    		\EIf {$H(p_e')> H(p_e)$} $p'_e \leftarrow p_e+ h\cdot stp$ \label{line_RBA:4}
    	\EndFor
    \Until {$|\hat{D}_{1}- D_{1}(\mathcal{G'})| \leq \tau $} 
 \end{algorithmic}
\end{algorithm}

\begin{figure}[h!]
    \begin{center}
        \centering
        \subfigure[backbone graph $G_b$]{
            \includegraphics[width=0.4\linewidth]{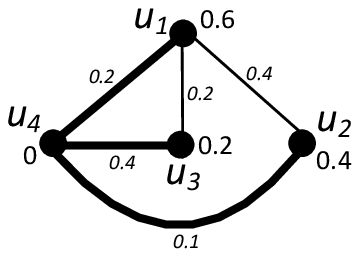}
        }
        \subfigure[output of \GDB]{
            \includegraphics[width=0.4\linewidth]{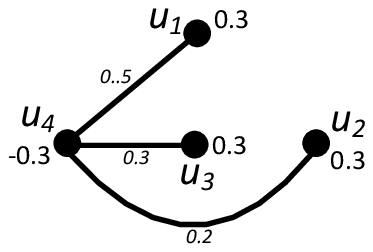}
        }
        \caption{\GDB\ example \label{fig:RBA}}
            \end{center}
\end{figure}

Figure \ref{fig:RBA} illustrates the execution of \GDB\ for $\delta^A$ with $\alpha=0.6$ and $h=1$ in the uncertain graph of Figure \ref{fig:RBA}(a). The edges of the backbone graph are depicted with bold, and the absolute discrepancies are shown next to each vertex. At each iteration, \GDB\ examines 
edges $(u_1,u_4)$, $(u_2,u_4)$, $(u_3,u_4)$ and decides their best probabilities. For example, for edge $(u_1,u_4)$, $p'_{(u_1,u_4)}=p_{(u_1,u_4)}+\frac{\delta(u_1)+\delta(u_4)}{2}=0.2+ \frac{0.6+0}{2}=0.5$. Figure \ref{fig:RBA}(b) contains the output of \GDB. Note that at this point, Equation (\ref{equ:rule}) cannot further modify the probability of any edge in the output graph. \GDB\ improved the objective function $D_1=\sum_{u \in V} \delta^2(u)$ from  to 0.56 to 0.36 and reduced the entropy from 3.85 to 2.60.  
\subsection{Expectation maximization degree}\label{seq:meth-3}
Since \GDB\ only updates the edge probabilities of the backbone graph $G_b=(V,E_b)$ (without inserting or removing edges), it is sensitive to the choice of $G_b$. On the other hand, \textit{Expectation-Maximization Degree} (\EMD) modifies both $E_b$ and the edge probabilities. \EMD\ is inspired by \textit{Expectation-Maximization} \cite{dempster1977maximum}, which is an iterative optimization framework that estimates two sets of interdependent unknown parameters. 
In our case, \EMD\ estimates the following sets of parameters: i) the set of edges in the sparsified graph and ii) their probabilities. 

Similarly to \GDB, the objective function is $D_1=\sum_{u \in V} \delta^2(u)$. \EMD\ starts with the input backbone graph, and the corresponding probabilities $p$ of $\mathcal{G}$. Then, it enters the iterative process, which consists of two phases. $E$-phase replaces edges of $E_b$ with edges from $E\setminus E_b$ considering the edge probabilities fixed. The new graph is denoted by $G'_b=(V,E'_b)$. $M$-phase calls \GDB\ to optimize the edge probabilities considering $G'_b=(V,E'_b)$ as fixed. We denote with $\hat{p}$ and $p'$ the probabilities of the previous and the current iteration respectively.

Equation (\ref{equ:rule}) provides a rule to estimate the values of $\mathbf{p'}$ with respect to some backbone graph $G_b$ and entropy parameter $h$. Accordingly, we need a rule to generate the graph $G'_b$ that minimizes the objective function with respect to a fixed set of probabilities $\mathbf{\hat{p}}$. To this end, we define the \textit{gain} of an edge $e=(u_0,v_0)$ as follows:
\begin{equation}\label{equ:gain}
	g(e)\big|_{p'_e} = \hat{\delta}^2(u_0)\big|_0-\hat{\delta}^2(u_0)\big|_{p'_{e}} + \hat{\delta}^2(v_0)\big|_0-\hat{\delta}^2(v_0)\big|_{p'_{e}}
\end{equation}	
where $p'_{e}$ is the probability of Equation (\ref{equ:rule}) and $\hat{\delta}(u_0)\big|_w$ is the degree discrepancy of vertex $u_0$ for $\hat{p}_{e}=w$. Intuitively, $g(e)$ quantifies the maximum improvement to $D_1$, incurred by including $e$ with the best probability $p'_e$. 

Our goal is to swap the edges of the current backbone $E_b$, with edges from $E\setminus E_b$ that have higher gain. An intuitive approach stores the edges of $E\setminus E_b$ in a dynamic max-heap $\mathcal{H}$ based on $g(e)$. At every iteration, it removes each edge $e=(u,v) \in E_b$ from $E_b$, and adds it in $\mathcal{H}$. Consequently, it recomputes the gains of all edges incident to $u$ or $v$ that have been affected by $e$'s removal, and updates $\mathcal{H}$ with the new gains. Then, it includes the top of $\mathcal{H}$, $e_H$, to the new backbone $E'_b$. Lastly, it updates $\mathcal{H}$ after recalculating the gain of edges incident to $e_H$. Unfortunately, this approach yields high computational cost for the following reasons:
\begin{myitemize}
\item The size of $\mathcal{H}$ is $O\big((1-\alpha)|E|\big)$. For small sparsification ratio, $\mathcal{H}$ is prohibitively large.
\item For every edge, $O(|E|/|V|)$ heap operations must be performed because each edge affects on average $2\cdot (1-\alpha)|E|/|V|$ other edges of $\mathcal{H}$. Thus, the total heap overhead of every $E$-phase is $O\big(\alpha(1-\alpha)|E|^2\log |V|/|V|\big)$.
\end{myitemize}
\EMD\ alleviates this overhead by maintaining a max-heap $\mathcal{H}_v$ of the vertices $V$, based on their discrepancy $\delta$. The method follows an approach similar to the above framework. This time however, an edge $e\in E_b$ is swapped with the edge $e_s$ that is incident to the top of $\mathcal{H}_v$, and has the highest gain. This greatly reduces the running time of \EMD\ compared to the above framework:
\begin{myitemize}
\item $\mathcal{H}_v$ has size $O(|V|)$, which is much smaller than $\mathcal{H}$.
\item For every edge, \EMD\ performs only $O(1)$ heap operations because each edge affects the discrepancy of only two vertices. Thus, at every $E$-phase, the total heap overhead is $O(\alpha|E|\log |V|)$.
\end{myitemize}

Algorithm \ref{algo:EMD} illustrates \EMD. Lines \ref{line_EMD:1}-\ref{line_EMD:2} initialize $\delta_A$, $E'$ and $\mathbf{p}$. 
Lines \ref{line_EMD:3}-\ref{line_EMD:4} contain $E$-phase. Initially, \EMD\ builds a max-heap $\mathcal{H}_v$ of the vertices $V$ based on $\delta_A$. At the iterative step, for each edge $e=(u,v)\in E_b$, \EMD\ excludes $e$ from $\E_b$, calculates its gain $g(e)$ and updates the corresponding entries of $\mathcal{H}_v$ for the affected vertices $u$ and $v$ (line 12). Then, it retrieves the top of $\mathcal{H}_v$, namely $v_H$, without removing it from the heap. For all edges $e_r\in E\setminus E_b$ adjacent to $v_H$, \EMD\ calculates their best probability according to Equation (\ref{equ:rule}). Using this probability, it computes the gain $g(e_r)$. Let $e_{max}=\arg \max \{g(e_r), g(e)\}$ be the edge with the maximum gain among edges $e_r$ and the original edge $e$. \EMD\ includes $e_{max}$ to the backbone graph and updates the incident vertices $u_{max},v_{max}$ in $\mathcal{H}_v$ (lines 19-20). This process repeats until all edges have been examined. Based on the new backbone graph, \textit{M}-phase further optimizes the probability assignment of $G'_b$ by calling \GDB (line \ref{line_EMD:5}). The procedure terminates when the improvement of an iteration is below a threshold $\tau$.  

\begin{algorithm}[ht]
  \caption{Expectation-Maximization Degree (\EMD)}
  \label{algo:EMD}
  \begin{algorithmic}[1]
    \Require uncertain graph $\mathcal{G}=(V,E,p)$, backbone graph $G_b=(V,E_b)$, entropy parameter $h$
    \Ensure sparse uncertain graph $\mathcal{G'}=(V,E',p')$ 
    \State $E' \gets \emptyset$ \label{line_EMD:1}
    \State initialize $\delta_A$ with expected degrees
	\ForAll {edge $e=(u,v)\in E_b$}	
		\State $ E' \gets E' \cup \{e\}$; $p'_e \gets p_e$
		\State $\delta_A(u)\gets \delta_A(u)-p_e$ \label{line_EMD:2}
	\EndFor	
	\Repeat	\label{line_EMD:3}
		\State $\hat{D}_{1} = D_{1}(\mathcal{G'})$ \Comment $E$-phase
		\State initialize max-heap $\mathcal{H}_v$ of vertices $V$ based on $|\delta_A|$ 
    	\ForAll {$e=(u,v)\in E'$}
			\State $\delta_A(u)\gets \delta_A(u)+p'_e$;$\delta_A(v)\gets \delta_A(v)+p'_e$	
			\State $E_b.\textsf{remove}(e)$; $p_e\gets 0$ 
			\State $\mathcal{H}_v.\textsf{update}(u,v)$;
			\State $v_H\gets \mathcal{H}_v.\textsf{top()}$
			\ForAll {edge $e_r\in E\setminus E_b$ adjacent to $v_H$}
				\State $w\gets$ probability of Equation (\ref{equ:rule})
				\State $g(e_r)|_w\gets$ gain of Equation (\ref{equ:gain})
			\EndFor
			\State $e_{max}=(u_{max},v_{max})\gets$ edge of max gain 
			\State $p_{max}\gets$ probability of $e_{max}$ 
			\State $E'_b.\textsf{add}(e_{max})$; $p_{e_{max}}\gets p_{max}$
			\State $\mathcal{H}_v.\textsf{update}(u_{max}, v_{max})$
		\EndFor \label{line_EMD:4}
		\State $\mathcal{G'}= \GDB(\mathcal{G},G'_b, h)$ \Comment $M$-phase  \label{line_EMD:5}
    \Until {$|\hat{D}_{1}- D_{1}(\mathcal{G'})| \leq \tau $} 
  \end{algorithmic}
\end{algorithm}

\begin{figure}[!h]
\vspace{-0.2cm}
\centering
\begin{tabular}{@{}c@{ \ }@{ \ }c@{}}
\includegraphics[width=0.21\textwidth]{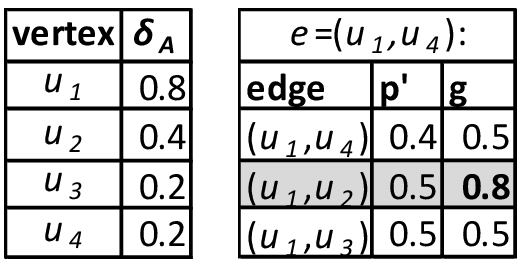} &
\includegraphics[width=0.2\textwidth]{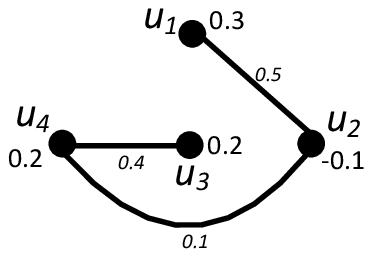} \\
(a) $\mathcal{H}_v$ and relevant edges  & (b) after first iteration  \\
at first iteration & of $E$-phase\\
\includegraphics[width=0.21\textwidth]{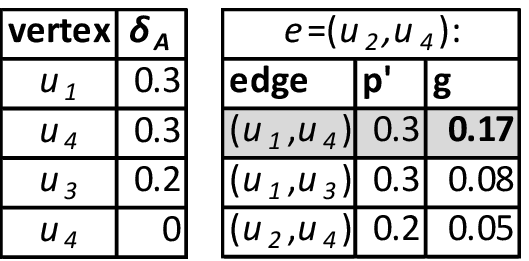} &
\includegraphics[width=0.2\textwidth]{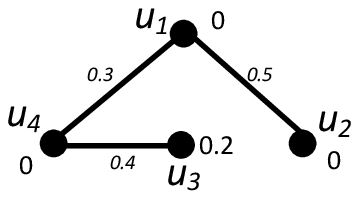} \\
(c) $\mathcal{H}_v$ and relevant edges  & (d) after second iteration \\
at second iteration & of $E$-phase\\

\end{tabular}
\caption{\label{fig:EMD} \EMD\ example}
\end{figure}

Figure \ref{fig:EMD} illustrates the execution of $E$-phase of \EMD, for $\delta^A$, in the uncertain graph of Figure \ref{fig:RBA}(a), with the same backbone $E_b$ (in bold) and entropy parameter $h=1$. At the iterative phase, edge $(u_1,u_4)$ is removed from $E_b$ and vertices $u_1,u_4$ update their discrepancy to the values of the left table \ref{fig:EMD}(a). The top of $\mathcal{H}_v$ is vertex $u_1$ and its adjacent edges are $(u_1,u_4), (u_1,u_2)$ and $(u_1,u_3)$. Equations (\ref{equ:rule}) and (\ref{equ:gain}) compute their best probability and gain respectively (right table of Figure \ref{fig:RBA}(a)). Edge $(u_1,u_2)$ has the highest gain 0.78, therefore it is included in the backbone $E'_b$. Figure \ref{fig:RBA}(b) demonstrates $E'_b$ with the corresponding probabilities (discrepancies) next to the edges (vertices). 

Let $(u_2,u_4)$ be the second edge to be examined. The left table of Figure \ref{fig:RBA}(c) shows $\mathcal{H}_v$ after the exclusion of $(u_2,u_4)$. Again, the top of $\mathcal{H}_v$ is vertex $u_1$. This time the relevant edges are $(u_1,u_4)$ and $(u_1,u_3)$, along with the edge under consideration $(u_2,u_4)$. The right table contains their corresponding best probability and gain. Edge $(u_1,u_4)$ has the highest gain, thus \EMD\ includes it to $E'_b$.   Figure \ref{fig:RBA}(d) contains $E'_b$. It can be easily verified that the remaining edge $(u_3,u_4)$ does not affect the backbone graph structure (it is removed and reinserted), thus the $E$-phase is complete. A subsequent $M$-phase on the updated backbone graph, calculates the new $\mathbf{p'}$ probabilities for the respective edges $(u_1,u_2) ,0.55$; $(u_1,u_4),0.2; $ and $(u_3,u_4),0.55$. Note that the resulting discrepancy $D_1=0.01$ has improved considerably. 
The original objective function $\Delta_1$ has also decreased to 0.2, from 1.2 in the backbone graph of Figure \ref{fig:RBA}(a). Similarly, entropy has decreased to 2.7 from 3.85 in the original graph.

\section{Preserving expected cuts}\label{sec:cuts}
\EMD\ cannot be applied for $k>1$ because our gain definition requires the computation of the discrepancy for all $k$-cuts that contain an edge, whose number is exponential. In the following we design a new rule that enables the application of \GDB\ to arbitrary values of $k\geq 1$. 
Since $|\delta(S)|$ is not differentiable, we again utilize the squared discrepancies $\delta^2(S)$, focusing on the absolute discrepancy $\delta_A$. Accordingly, the objective function is:
\vspace {-0.1in}
\begin{equation*}
D_k=\sum_{i=1}^k\sum_{\substack{S \subseteq V ,|S|=k}} \delta_A^2(S) 
\end{equation*}
Its derivative with respect to $p'_{e}$ for an edge $e=(u_0,v_0)$ is:
\vspace {-0.1in}
\begin{align*}\label{equ:cutsderivative}
\frac{\partial D'_k}{\partial p'_{e}}=-2 \sum_{i=1}^k\sum_{\substack{S \subseteq V, |S|=k \\ u_0 \in S, v_0 \notin S}} \delta_A'(S)- 2 \sum_{i=1}^k\sum_{\substack{S \subseteq V, |S|=k \\ v_0 \in S, u_0 \notin S}} \delta_A'(S)
\end{align*}

Changing the probability of edge $e$ from $\hat{p}_e$ to $p'_e$ the discrepancies of all cuts that contain $e$ become:  
\begin{equation}\label{equ:deltastarA}
\delta_A'(S)= \hat{\delta}_A(S)+\hat{p}_e-p'_e
\end{equation}
Setting the derivative equal to zero and solving with respect to $p'_{e}$, using Equation (\ref{equ:deltastarA}), we obtain the best probability for $e$:
\vspace {-0.1in}
\begin{align}\label{equ:sumrule}
p'_{e}=\hat{p}_{e}+\frac{\sum\limits_{i=1}^k\Big(\sum\limits_{\substack{S \subseteq V, |S|=k \\ u_0 \in S, v_0 \notin S}} \hat{\delta}_A(S)+\sum\limits_{\substack{S \subseteq V, |S|=k \\ v_0 \in S, u_0 \notin S}} \hat{\delta}_A(S)\Big)}{\sum\limits_{i=1}^k\Big(\sum\limits_{\substack{S \subseteq V, |S|=k \\ u_0 \in S, v_0 \notin S}}1+\sum\limits_{\substack{S \subseteq V, |S|=k \\ v_0 \in S, u_0 \notin S}}1\Big)}
\end{align}

The computation of the above equation is not tractable due to the fact that we need to enumerate all $k$-cuts that contain edge $e=(u_0,v_0)$. To avoid this we introduce the following enumeration function:
\vspace {-0.1in}
\begin{equation*}
\binom{n}{k}_{\Sigma} = \left\{
  \begin{array}{lr}
  	0    & \text{if } k < 0 \\
   \sum\limits_{i=0}^k\binom{n}{i} & \text{if } k> 0 
    
  \end{array}
  \right.
\end{equation*}

We count how many times the discrepancy of an edge is present in the sum of the numerator. If the edge is incident to $u_0$ or $v_0$, it will be counted $\binom{n-3}{k-1}_{\Sigma}$ times because we restrict its other vertex from entering $S$. All other edges, not incident to $u_0$ or $v_0$, will be counted $2\binom{n-4}{k-2}_{\Sigma}$ times in the sum, since only one of their incident vertices belongs to $S$. 
\vspace {-0.1in}
\begin{align*}
\sum\limits_{i=1}^k\sum\limits_{\substack{S \subseteq V, |S|=k \\ v_0 \in S, u_0 \notin S}}  \hat{\delta}_A(S) = \binom{n-3}{k-1}_{\Sigma} \hat{\delta}_A(u_0) +2\binom{n-4}{k-2}_{\Sigma} \hat{\Delta}(e)
\end{align*}
\vspace{-0.15in}
where:	
\vspace {-0.1in}
\begin{align*}
\hat{\Delta}(e)=\sum_{\mathclap{\substack{(u_1,v_1) \in E, \\ u_1 \neq u_0, v_1 \neq v_0}}} p_{u_1v_1}-\hat{p}_{u_1v_1}
\end{align*}
Therefore Equation (\ref{equ:sumrule}) reduces to:
\begin{align}\label{equ:cutrule}
p'_{e}=\hat{p}_{e}+\frac{ \binom{n-3}{k-1}_{\Sigma} \Big(\hat{\delta}_A(u_0)+\hat{\delta}_A(v_0)\Big)+4\binom{n-4}{k-2}_{\Sigma} \hat{\Delta}(e)}{2\binom{n-2}{k-1}_{\Sigma} }
\end{align}

Equation (\ref{equ:cutrule}) proposes a probability change that weights the degree discrepancies $\hat{\delta}(u_0)$ and $\hat{\delta}(v_0)$ versus the discrepancy of the edges that are not incident to $u_0$ and $v_0$, $\hat{\Delta}(e)$. The best probability $p'_{e}$ can exceed [0,1]. However, since the objective function is again convex (i.e.,$\frac{\partial^2 D_k}{\partial {p_e}^2}>0$), the optimal probability is:
\vspace {-0.09in}
\begin{align}\label{general_rule}
p'_{e}=\tensor*[_0]{\Bigg\lfloor}{} \hat{p}_{e}+h\frac{ \binom{n-3}{k-1}_{\Sigma} \Big(\hat{\delta}_A(u_0)+\hat{\delta}_A(v_0)\Big)+4\binom{n-4}{k-2}_{\Sigma} \hat{\Delta}(e)}{2\binom{n-2}{k-1}_{\Sigma} }\Bigg\rceil^1
\end{align}
where the entropy parameter $h\in [0,1]$ tunes the step size of gradient decent. 

The only modification of \GDB, is that in line 7 of Algorithm \ref{algo:RBA}, the optimal step \textit{stp} is replaced by the ratio in Equation (\ref{equ:cutrule}). Special cases of the general rule include $k=1$, $k=2$ and $k=n$.
For $k=1$ and absolute discrepancies $\delta_A$, the above equation reduces to Equation (\ref{equ:rule}), and thus takes into account only the degree discrepancies. For $k=2$, the best probability is:
\vspace {-0.09in}
\begin{align}\label{equ:2cutrule}
p^{2}{'}_{e}=\tensor*[_0]{\Bigg\lfloor}{}\hat{p}_{e}+h\frac{ (n-2) \Big(\hat{\delta}_A(u_0)+\hat{\delta}_A(v_0)\Big)+4\hat{\Delta}^e}{(2n-2) }\Bigg\rceil^1
\end{align}

For $k=n$, the update rule changes to the following formula, which distributes the cumulative probability of eliminated edges to all the remaining ones. This corresponds to random probability reassignment.
\vspace {-0.1in}
\begin{align}\label{equ:ncutrule}
p^{n}{'}_{e}=\tensor*[_0]{\Bigg\lfloor}{}\hat{p}_{e}+h\smashoperator{\sum_{e_1\in E'\setminus \{e\}}}(p_{e_1}-\hat{p}_{e_1})\Bigg\rceil^1
\end{align}

The general rule of Equation (\ref{general_rule}) is analytic and does not require any enumeration of cuts. Consequently, the running time of \GDB\ is insensitive to $k$ and depends only on the convergence speed of gradient descent. 
\section{Experiments} \label{sec:exp}
\vspace{-1mm}
In our evaluation, we use two real undirected uncertain graphs with various sizes, densities, and edge probabilities, summarized in Table \ref{tab:datasets}. \flickr\ \cite{PBGK10} is a social network, where edge probabilities are based on the principle that similar interests indicate social ties. This is the densest dataset; a vertex has on average about 130 neighbours. \twitter~\cite{bonchi2014core} is extracted from the popular online micro-blogging service. Probabilities denote the influence that the associated users exert on each other. Although sparser than \flickr, \twitter\ has higher average probability on the edges.

\begin{table} [h]
\small
\centering
{%
\begin{tabular} { c|c|c|c|c|c}
\hline    
{\textbf{dataset}} &
{\textbf{vertices}}&
{\textbf{edges}} &
{\textbf{$|E|/|V|$}} & 
{$\mathbb{E}[p_e]$}&
{{$\mathbb{E}[d_u]$}} \\ \hline
\flickr	&	$78\,322$	&	$10\,171\,509$	&	$129.89$	&	$0.09$	&	$22.93$ \\
\twitter	&	$26\,362$	&	$663\,766$	&	$25.17$	&	$0.15$	&	$7.71$ 		\\
 \hline
\multirow{4}{*}{\textsf{Synthetic}}& \multirow{4}{*}{$1\,000$}
&$77\,099$ & $77.1$ &\multirow{4}{*}{0.09}&12.7\\
& & $147\,565$ & $147.5$ & &24.3\\
& & $269\,325$ & $269.3$ & &44.3\\
& & $435\,336$ & $435.3$ & &71.5 \\
\hline
  \end{tabular}
}
\vspace{+3mm}
\caption{Characteristics of datasets}
\label{tab:datasets}
\vspace{-5mm}
\end{table}

%
%
%
%

In order to asses the behaviour of the methods in graphs with increasing density, we also use 4 synthetic undirected datasets, whose characteristics are summarized in Table \ref{tab:datasets}. They all stem from an induced subgraph of Flickr with 1000 vertices, where edges have been added between random pairs of vertices, until the density becomes 15, 30, 50, 90 \% of the complete graph. The additional edge probabilities follow the same distribution as the original Flickr.  

All methods were implemented in C++ and executed in a single core of an Intel Xeon E5-2660 with 2.20GHz {\sc cpu} and 96GB {\sc ram}. Section \ref{sec:expOur} assesses variants of the proposed methods on the objective function of Problem 1 for various values of $k$ in order to identify the best ones depending on the problem characteristics. We refer to the graph characteristics preserved by these objective functions (degrees/cut sizes and entropy) as structural properties. Section \ref{sec:expBench} and \ref{sec:expQueriesBench} compare representative variants against the benchmarks on structural properties and common graph queries, respectively. 
\subsection{Assessment of proposed techniques}\label{sec:expOur}

We evaluate the proposed methods \GDB\ (\textit{Gradient Descent Backbone}) and \EMD\ (\textit{Expectation Maximization Degree}) on structural properties using absolute degree discrepancy\footnote{The respective $\delta^R$ results are similar and omitted.}. As a benchmark, we use \LP, the \textit{linear programming} technique of Section \ref{seq:meth-1}, which, given a backbone graph, yields the optimal probability assignment that minimizes $\Delta_1$ with entropy parameter $h=0$. Because \LP\ fails to terminate within reasonable time in the real datasets, the experiments are performed on an induced subgraph of Flickr that consists of 5,000 vertices and 655,275 edges, and was extracted using Forest Fire \cite{Leskovec2006}.

We use the following notation to differentiate among variants of each method:
\begin{myitemize}
\item $A$ and $R$ superscripts denote the variants that aim at minimization of the absolute $\delta_A$ and relative $\delta_R$ discrepancy, respectively. The first type favors nodes with high degree by targeting the absolute error, whereas the second treats all degrees equally.
\item $t$ suffix signifies that the backbone graph is generated by Algorithm \ref{algo:BGG}, which ensures connectivity. Absence of this suffix means that the backbone is created by Monte Carlo sampling on the original graph, until reaching of $\alpha |E|$ edges. We refer to this as the \textit{random backbone}.
\item $k=\{2,n\}$ subscript denotes \GDB\ preserving $k$-cuts. Absence of this subscript implies that $k=1$ (i.e., preservation of the expected degrees). 
\end{myitemize}

Table \ref{tab:ours} shows the mean absolute error (MAE) of the absolute degree discrepancy $\delta_A$ of variants of \GDB, \EMD\ and \LP\ for sparsification ratio $\alpha$ ranging from $8\%$ to $64\%$.
$GDB^A_n$ has by far the worst performance as it randomly distributes the missing probabilities to all edges of $\mathcal{G'}$. The accuracy of $\GDB^A$, $\GDB^R$ and $\GDB^A_2$ is similar and comparable to \LP\ for the corresponding backbone graph. The backbone graphs generated by Algorithm \ref{algo:BGG} benefit all variants (compared to random backbones). However, for $\alpha=8\%$, the spanning nature of the graph increases the discrepancy of some vertices that would be otherwise disconnected. In this case, \LP\ and \GDB\ perform better using random backbones as input. In general, \EMD\ improves the accuracy compared to the respective $\GDB$ versions by re-structuring the backbone. Methods that preserve the relative discrepancy have similar behaviour to those aiming at absolute discrepancy. The variant with the best overall performance is $\EMD^R$-$t$, which achieves the highest accuracy for all values of sparsification ratio $\alpha>8\%$, shown in bold in Table \ref{tab:ours}. 

\begin{table} [h!]
\small
\centering
{%
\begin{tabular} { c|c|c|c|c}
\hline    
{}	& {8\%} & {16\%} & {32\%} & {64\%} \\ \hline

$\LP$	&\textbf{1.04} &	2.56E-02	& 4.22E-03	& 2.70E-04 \\
$	\GDB^A$	&	1.19 &	2.68E-02	&	4.38E-03	&	3.92E-04 \\
$	\GDB^R$	&	1.21	&	2.67E-02	&	4.38E-03	&	3.92E-04 \\
$	\GDB^A_2$	&	1.73	&	4.25E-02	&	4.74E-03	&	6.04E-04 \\
$	\GDB^A_n$	&	5.78	&	3.27	&	2.17	&	1.8 \\
\hline
$\EMD^A$	&	1.33	&	2.56E-02	&	1.22E-03	& 7.89E-13 \\
$\EMD^R$	&	1.35	&	2.18E-02	&	1.43E-03	&	1.79E-12 \\

\hline
$	\LP$-$t$	&	2.27	&	2.95E-04	&	2.99E-05	&	2.62E-12 \\
$	\GDB^A$-$t$	&	3.54	&	1.78E-03	&	1.82E-04	&	1.66E-04 \\
$	\GDB^R$-$t$	&		2.47	&	4.11E-04	&	2.99E-05	&	2.62E-12 \\

\hline
$	\EMD^A$-$t$	&	2.53	&	1.83E-04	&	4.81E-12	&	8.23E-13 \\
$	\EMD^R$-$t$&		2.55	&	\textbf{9.23E-05}	&	\textbf{8.17E-13}	&	\textbf{7.34E-13} \\
\hline

  \end{tabular}
}
\vspace{2mm}
\caption{Mean Absolute Error (MAE) of absolute degree discrepancy $\delta_A(u)$ (Flickr reduced)}
\label{tab:ours}
 \vspace{-1mm}
\end{table}

\begin{figure}[!h]
    \begin{center}
        \centering
       \hspace*{-0.5cm}
        \subfigure[MAE of $\delta_A(S)$]{
            \includegraphics[width=0.49\linewidth]{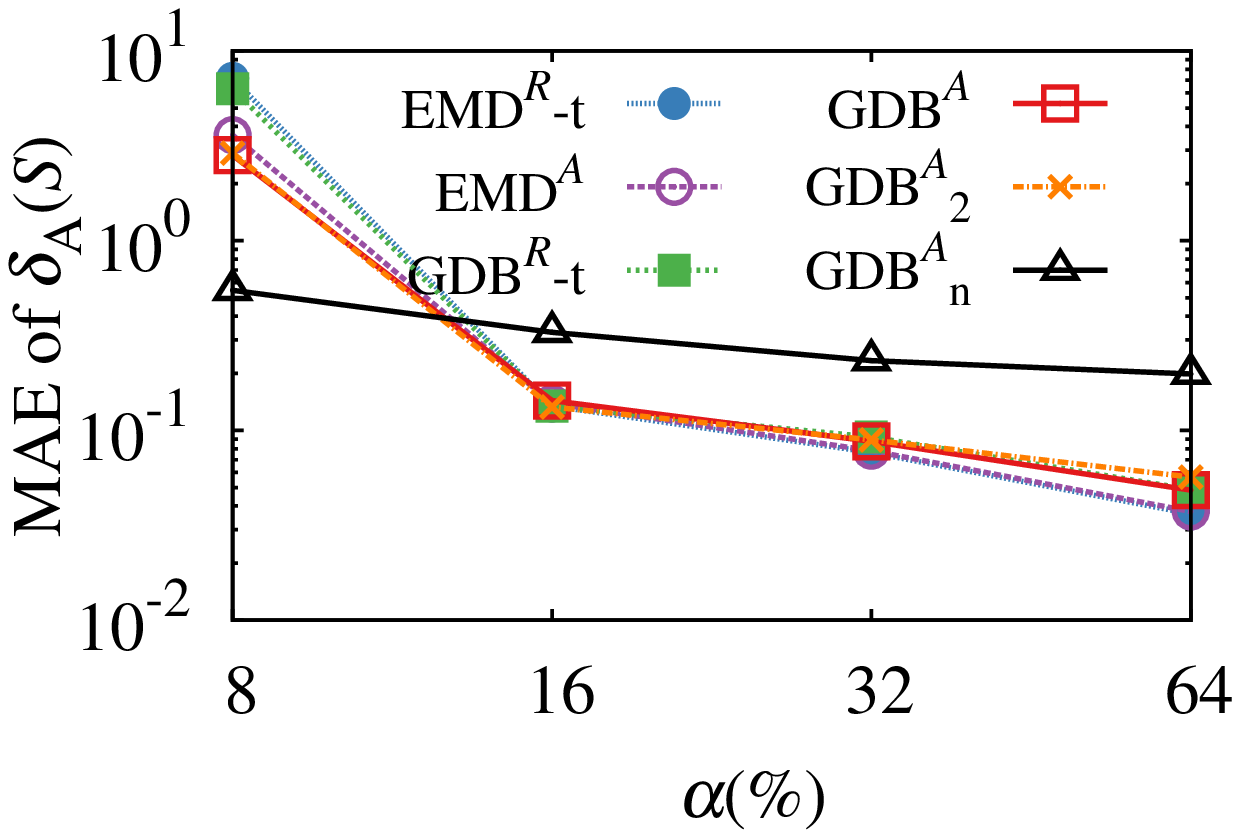}
        }
        \subfigure[Execution time]{
            \includegraphics[width=0.49\linewidth]{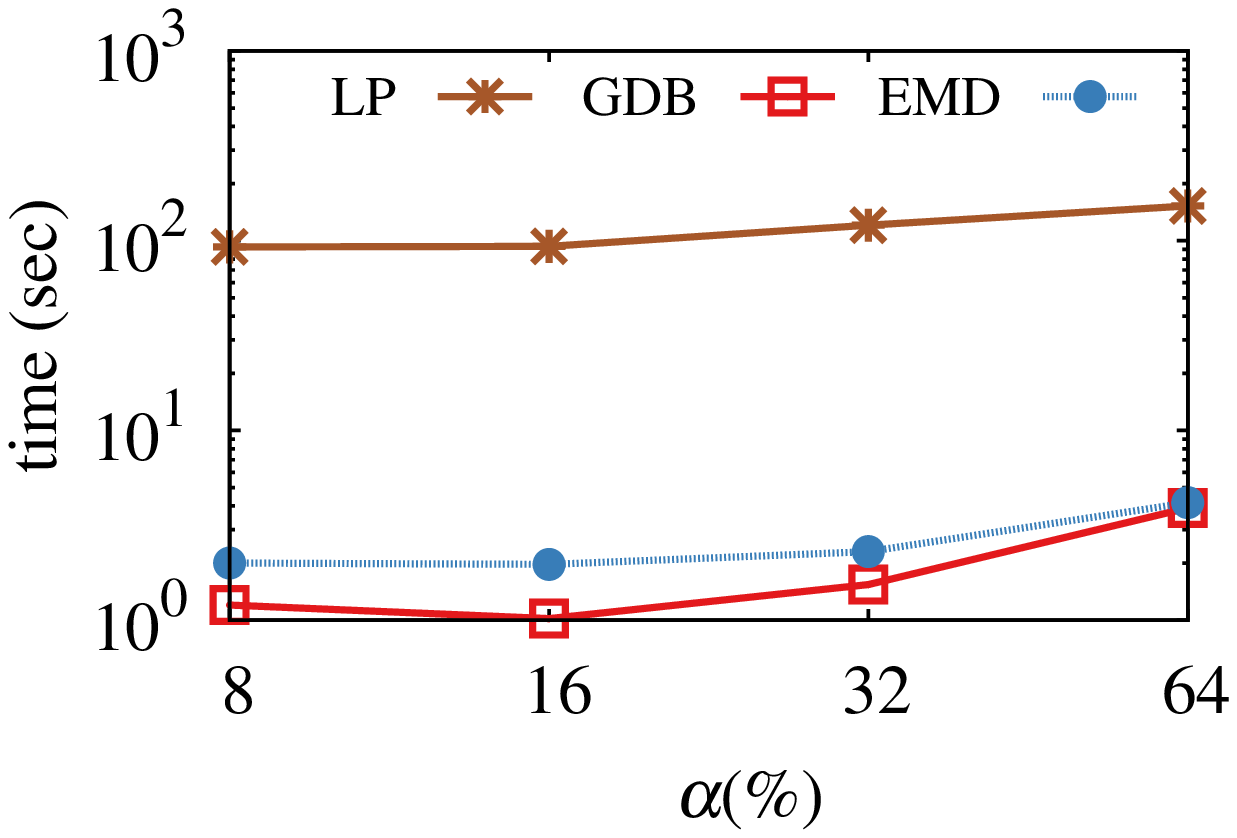}
        }
        \caption{(a) MAE of cut size discrepancy $\delta_A(S)$ and (b) execution time (Flickr reduced)}
        \label{fig:ours}
 \vspace{-0.5cm}
    \end{center}
\end{figure}

Figure \ref{fig:ours}(a) shows the MAE of the cut discrepancies versus $\alpha$. Since it is intractable to measure all cuts, we randomly select 1000 $k$-cuts for $k=1$ up to $|V|$ for each value of $k$, and we compute the average absolute discrepancy. \LP\ is excluded because it explicitly aims at $\Delta_1$. Similar to Table \ref{tab:ours} and for the same reasons $\GDB^A_n$ under-performs the other variants for $\alpha>8$. The superior performance of $\GDB^A_n$ for $\alpha=8\%$ is explained as follows. In Flickr (also in the reduced graph), the average probability is 0.09; thus, the expected number of edges is approximately $0.09|E|$. For $\alpha<9\%$ the sparsified graph does not contain enough edges to reach the expected number under any probability assignment. $\GDB^A_n$ assigns the maximum probability $p=1$ to all available edges. On the other hand, the rest of the methods, respecting the constraints on degree and 2-cuts do not entirely redistribute the missing probabilities. For $\alpha>8\%$, this ceases to be the case and they yield high accuracy. 

Figure \ref{fig:ours}(b) illustrates the running time as a function of $\alpha$. 
For both \GDB\ and \EMD\ the running time is independent of the discrepancy (absolute or relative), and the structure of the backbone graph. In addition, \GDB\ optimizes cut sizes using analytic equations; thus its performance is insensitive to $k$. Accordingly, the plots of \GDB\ and \EMD\ capture all variants of each method. Both techniques are significantly faster than \LP, which cannot be applied for large graphs. \EMD\ is slower than \GDB\ since it invokes it as a module for the $M$-phase. However, the overhead is small, which confirms the efficiency of the vertex (as opposed to edge) heap.

In order to compare against the benchmark methods, we select as representative variants of our techniques 
$\EMD^R$-$t$ and $\GDB^A$. $\EMD^R$-$t$ has the most balanced performance for the settings of Table \ref{tab:ours} and Figure \ref{fig:ours}(a). $\GDB^A$ is in general inferior, but it outperforms $\EMD^R$-$t$ for sparsification ratio $\alpha=8\%$. Moreover, the two variants collectively cover all combinations of discrepancy type and backbone structure. Thus, in the following, the terms \EMD\ and \GDB\ refer to these variants. Since in the remaining we use the real graphs, \LP\ is excluded due its high cost.

\begin{figure}[!h]
    \begin{center}
        \centering
       \hspace*{-0.5cm}
        \subfigure[MAE of $\delta_A(S)$]{
            \includegraphics[width=0.49\linewidth]{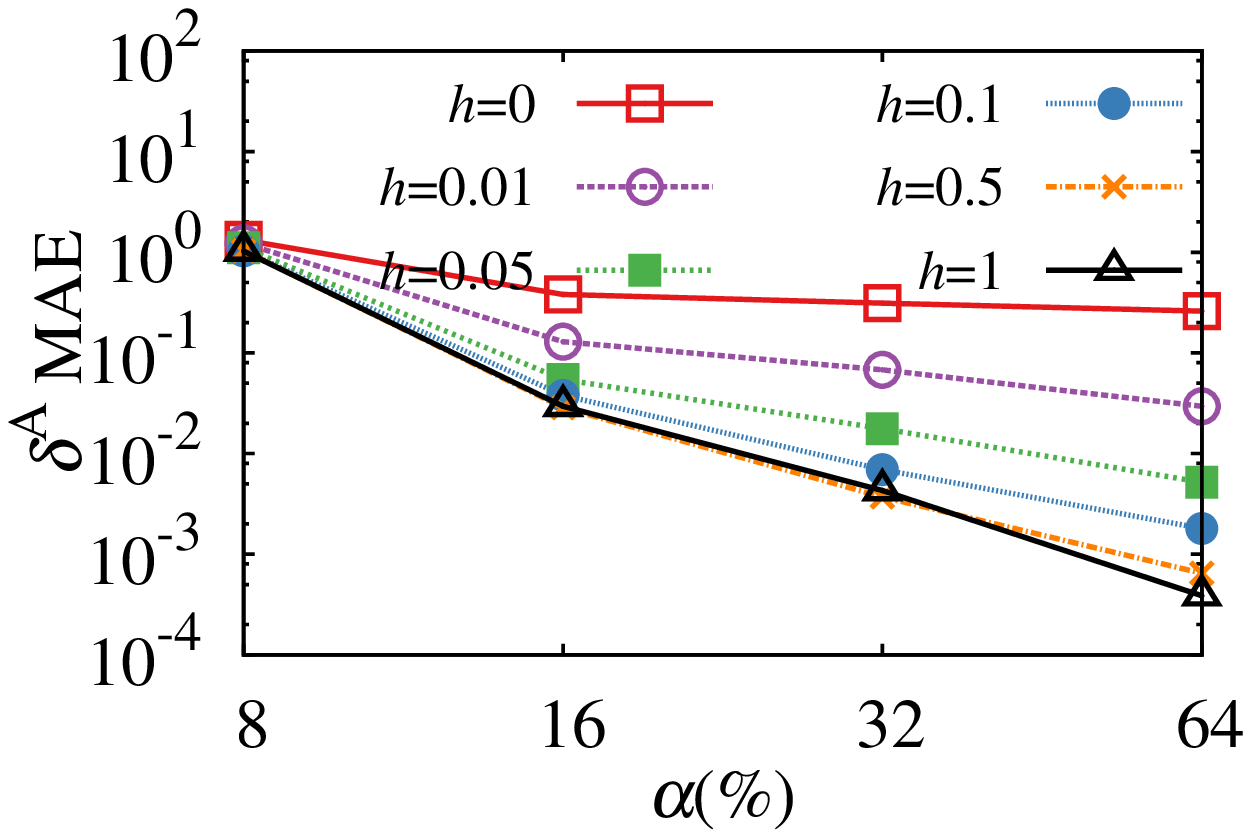}
        }
        \subfigure[Relative entropy $\frac{H(\mathcal{G'})}{H(\mathcal{G})}$]{
            \includegraphics[width=0.49\linewidth]{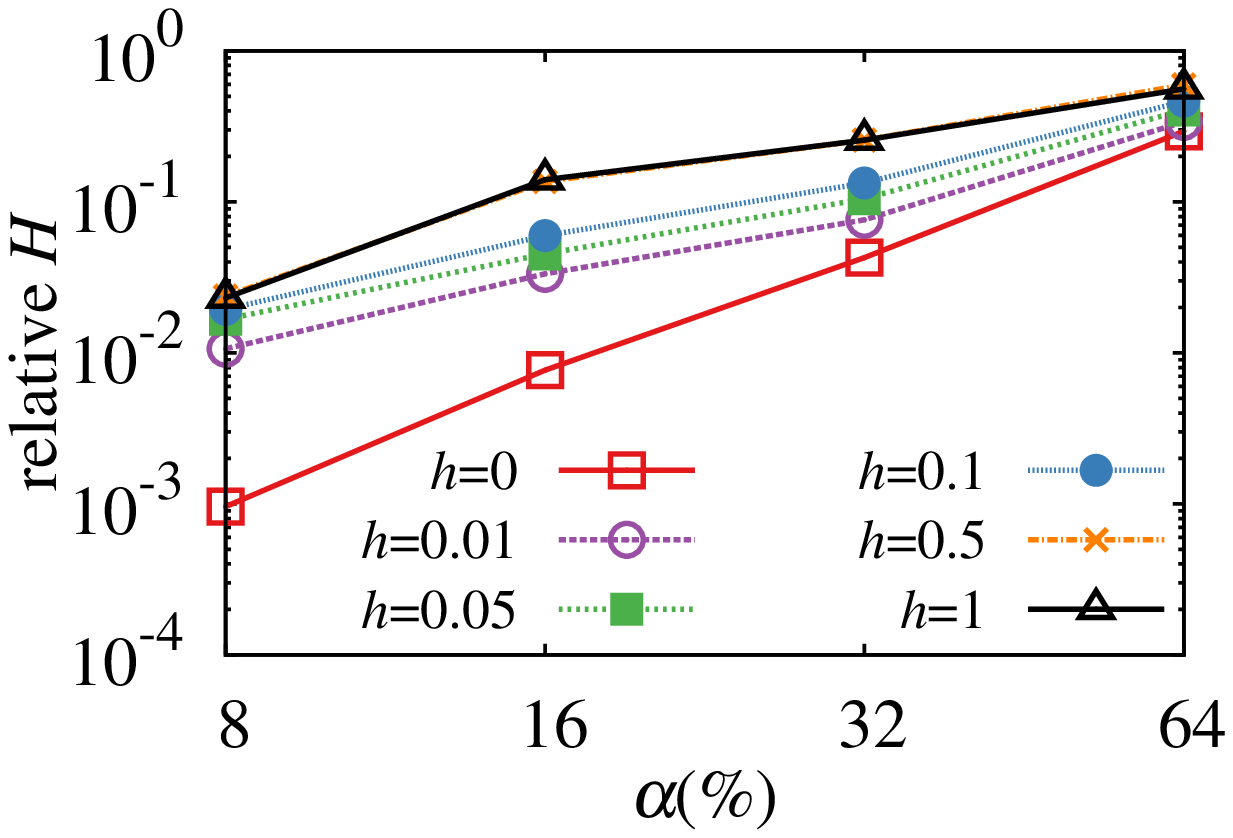}
        }
        \caption{Effect of entropy parameter $h$ on \GDB\ (Flickr reduced)}
        \label{fig:ours_entropy}
 \vspace{-0.5cm}
    \end{center}
\end{figure}

A final remark concerns the fine-tuning of entropy parameter $h\in [0,1]$ in \GDB\ (and \EMD, since \GDB\ constitutes one of its modules). Recall from Section \ref{seq:meth-2} that $h$ reduces the gradient descent step size when the optimal probability assignment increases entropy. Figure
\ref{fig:ours_entropy}(a) plots the MAE of the absolute degree discrepancy $\delta_A$ versus the sparsification ratio, for various values of $h$. In the extreme case of $h=0$, \GDB\ yields poor performance for $\delta_A$ because it discards \textit{any} probability assignment that increases the edge entropy. On the other hand, for $h=1$ \GDB\ yields the best result on $\delta_A$, but the worst entropy values as shown in Figure \ref{fig:ours_entropy}(b), which plots the relative entropy $\frac{H(\mathcal{G'})}{H(\mathcal{G})}$ versus $\alpha$.
 Intermediate values of $h$ span between these two extremes. In the remaining, we set $h=0.05$ as it has the most balanced performance.

\subsection{Comparison with benchmarks on structural properties}\label{sec:expBench} 
  
\begin{figure}[h!]
    \begin{center}
        \centering
        
        \hspace*{-0.45cm}
        \subfigure[MAE of $\delta_A (u)$ (Flickr)]{
            \includegraphics[width=0.51\linewidth]{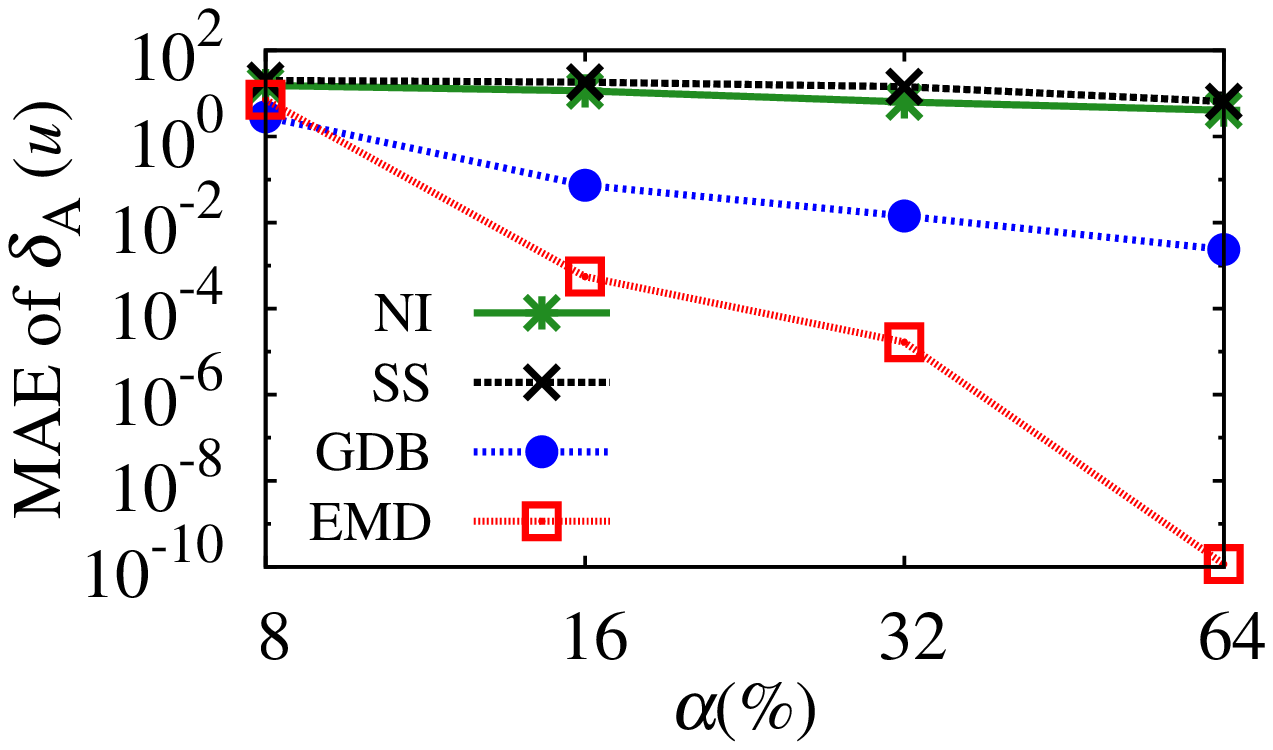}
        }\vspace{-2mm}
       \hspace*{-0.45cm}
        \subfigure[MAE of $\delta_A (S)$ (Flickr)]{
            \includegraphics[width=0.51\linewidth]{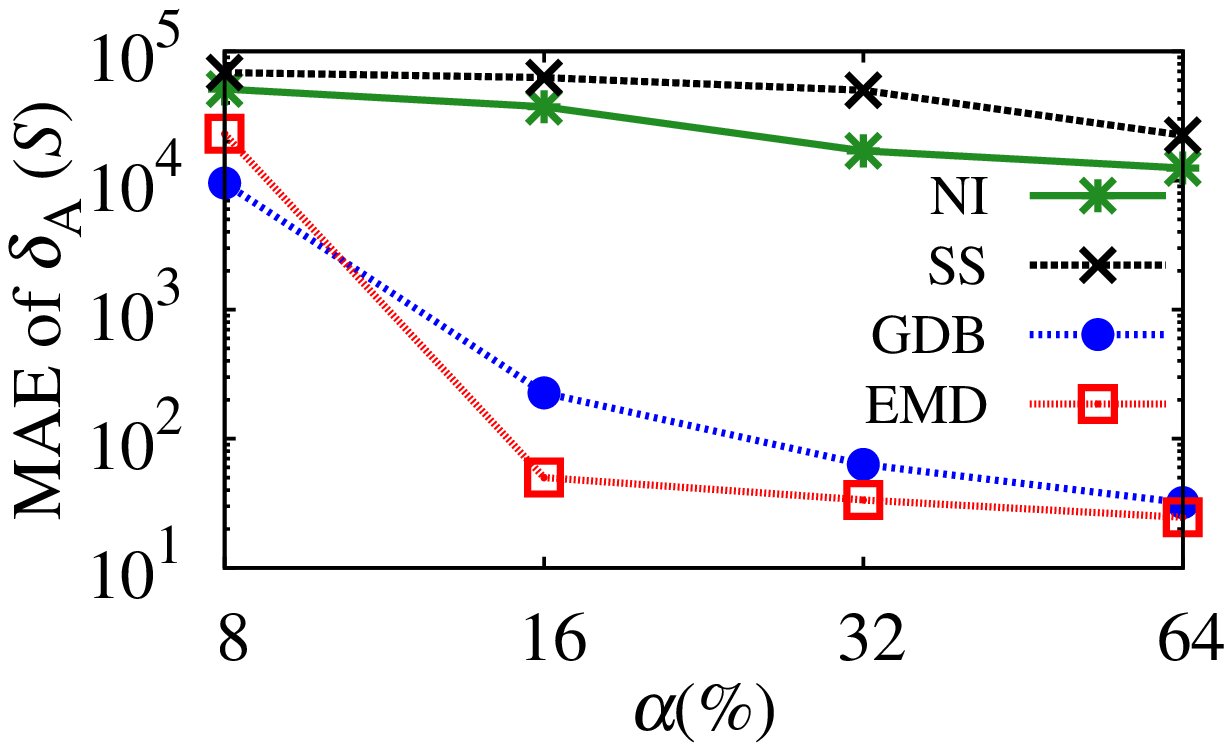}
        }
        \vspace{-1mm}
       \hspace*{-0.45cm}
        \subfigure[MAE of $\delta_A(u)$ (Twitter)]{
            \includegraphics[width=0.51\linewidth]{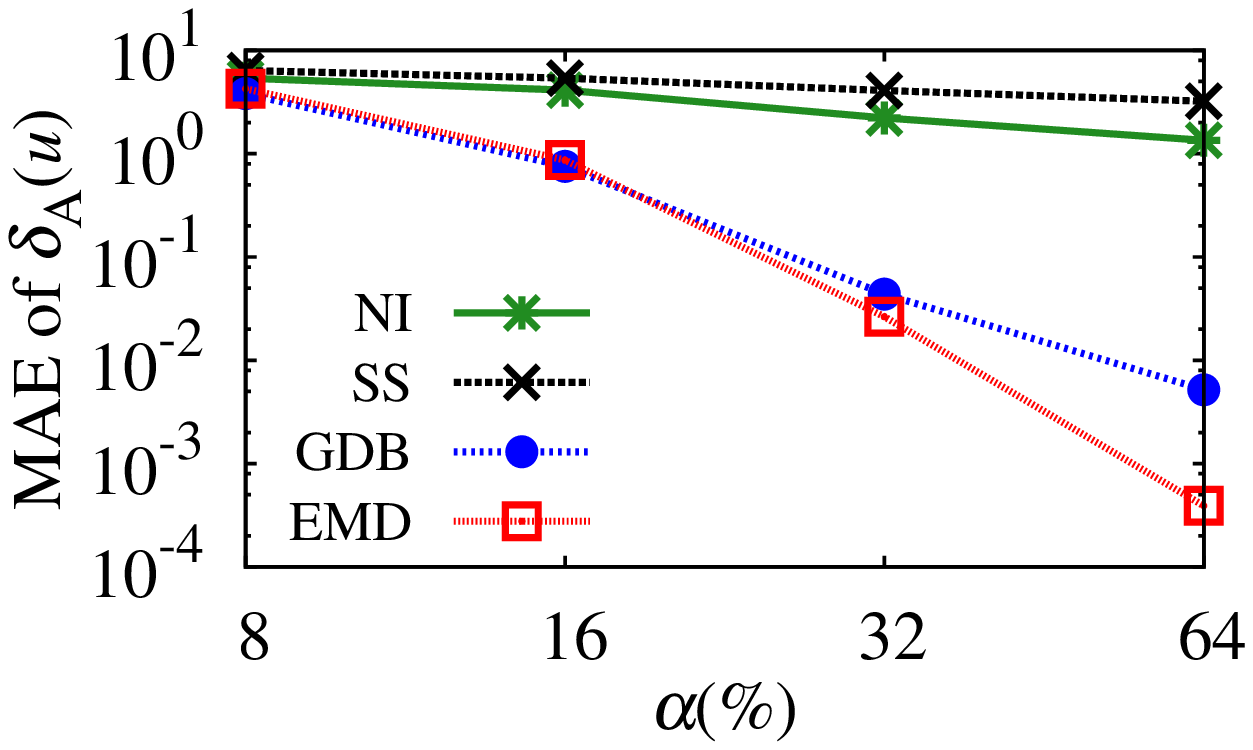}
        }\vspace{-1mm}
       \hspace*{-0.45cm}
        \subfigure[MAE of $\delta_A (S)$ (Twitter)]{
            \includegraphics[width=0.51\linewidth]{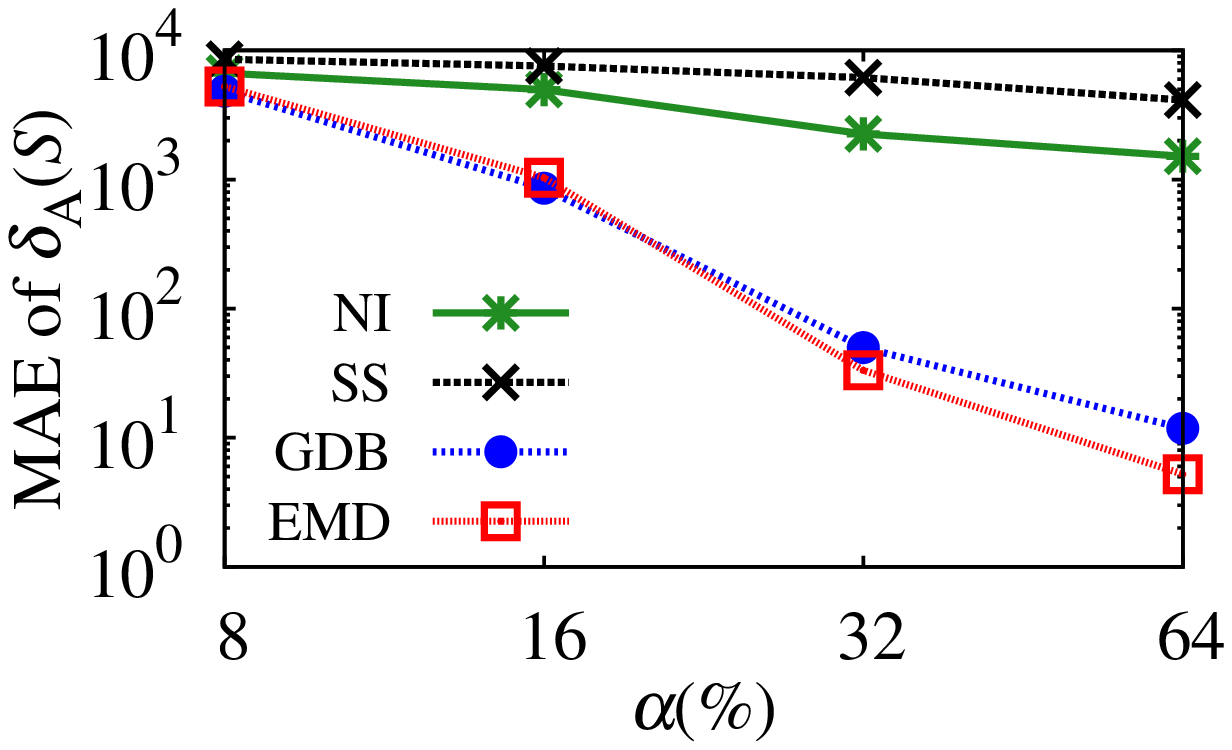}
        }
                
        \caption{MAE of absolute degree discrepancy $\delta_A(u)$ and absolute cut size discrepancy $\delta_A(S)$ (real datasets)}
        \label{plot:objective}
\vspace{-0.5cm}
    \end{center}
\end{figure}

We compare \EMD\ and \GDB\ against the benchmarks \NI\ and \SS. Recall that \NI\ constitutes the adaptation of a cut-based deterministic sparsification method, whereas \SS\ extends a spanner-based technique to the uncertain setting. 
Figure \ref{plot:objective} plots the MAE of the absolute degree discrepancy $\delta_A(u)$ and the MAE of the cut discrepancy $\delta_A(S)$ (objective function for $k=1$ and $k\geq 1$, respectively) versus $\alpha$. The proposed methods consistently outperform the benchmarks for both structural properties in all datasets. The low accuracy of \SS\ can be explained by the fact that it was designed to capture shortest path distances, instead of cuts or degrees. \NI\ is comparable to the proposed techniques for small values of $\alpha$ in Twitter that have high edge probabilities. In these cases, the backbone graph is almost deterministic (most edges have probability 1) and there is little space for improvement by  \EMD\ and \GDB. For the other settings, \NI\ fails because it assumes unbounded weights. Bounding the maximum weight to 1 seriously affects both its performance and its theoretical guarantees: \NI\ yields a mild probability redistribution that fails to preserve degrees and cuts in practice. Moreover, \NI\ is designed for dense graphs with $E=\Theta(n^2)$, whereas the evaluated datasets are much sparser. As expected from Table \ref{tab:ours} and Figure \ref{fig:ours}(a), in most settings \EMD\ outperforms \GDB.

\begin{figure}[h!]
    \begin{center}
        \centering
        
        \hspace*{-0.45cm}
        \subfigure[MAE of $\delta_A (u)$ (Synthetic)]{
            \includegraphics[width=0.51\linewidth]{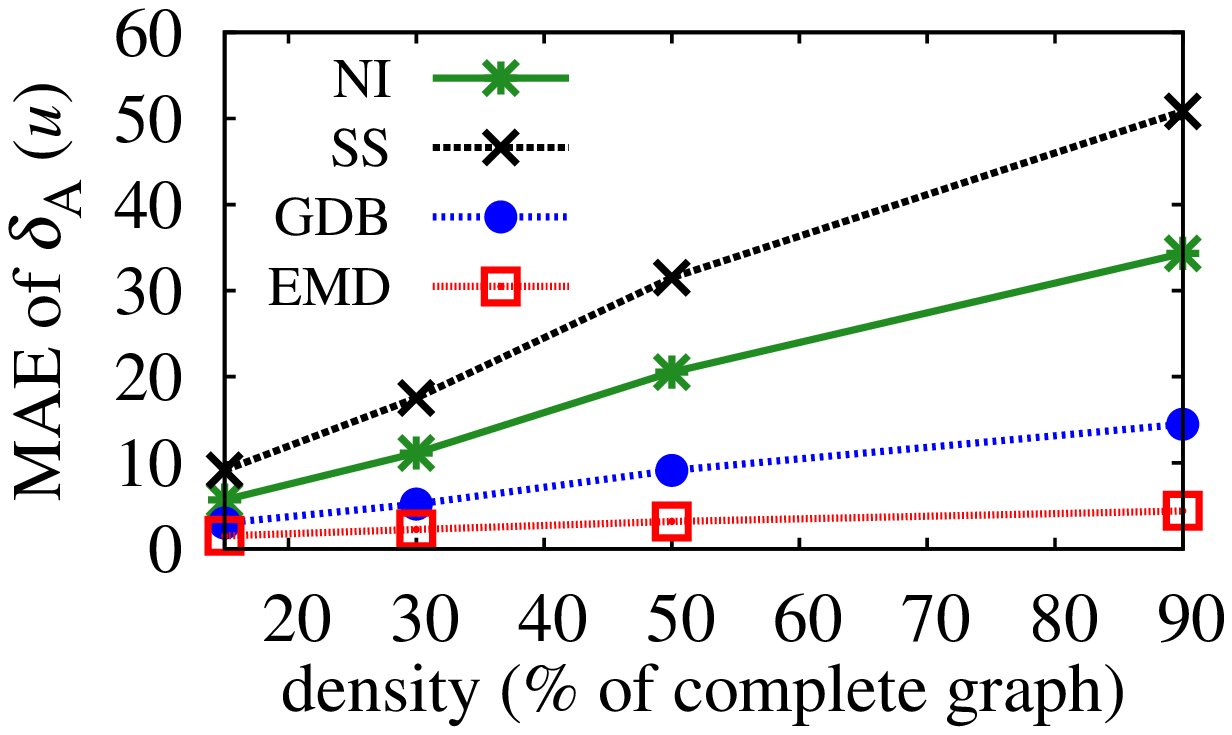}
        }
       \hspace*{-0.45cm}
        \subfigure[MAE of $\delta_A (S)$ (Synthetic)]{
            \includegraphics[width=0.51\linewidth]{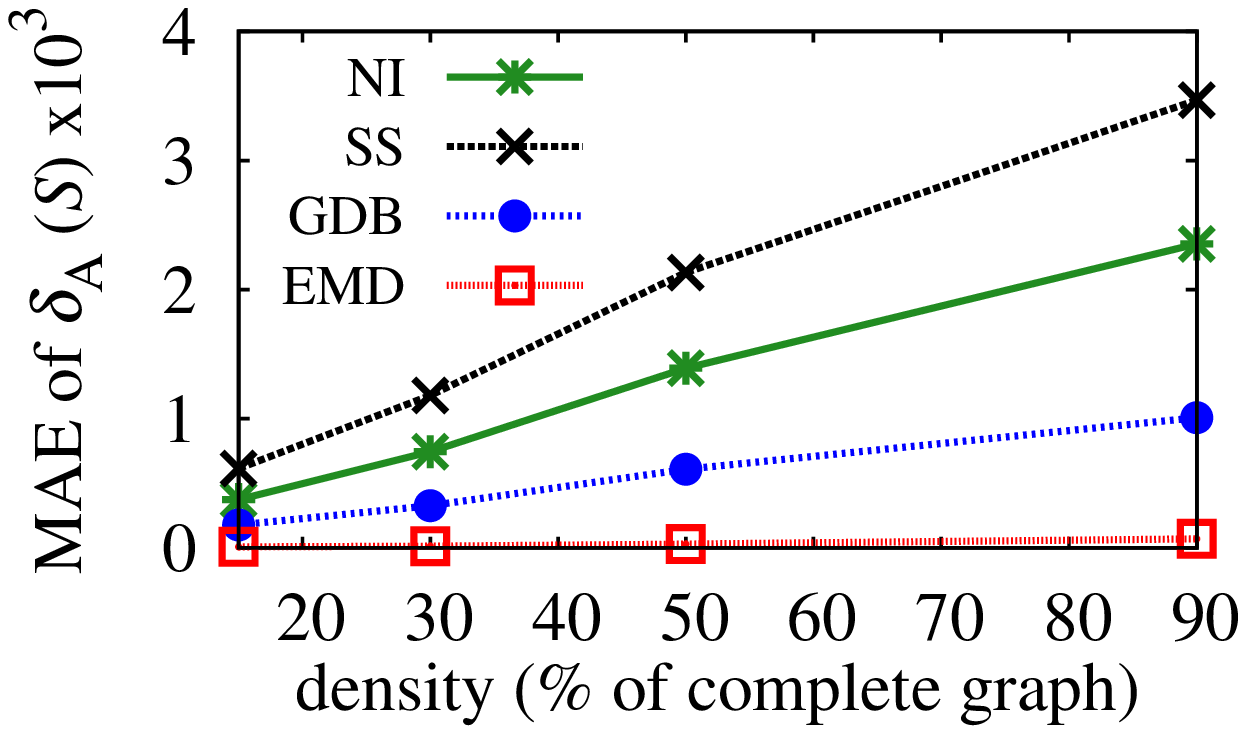}
        }
                
        \caption{MAE of absolute degree discrepancy $\delta_A(u)$ and absolute cut size discrepancy $\delta_A(S)$ versus graph density (\% of complete graph) (synthetic datasets)}
        \label{plot:objective_dense}
 \vspace{-0.3cm}
    \end{center}
\end{figure}

Figure \ref{plot:objective_dense} plots MAE of $\delta_A(u)$ and $\delta_A(S)$  as a function of the graph density (percentage of the complete graph), on the synthetic datasets, which are denser than the real ones. The sparsification ratio $\alpha$ is fixed to 16\%. As the graph density increases, all methods yield increasing error. To see why this happens, consider \SS\ that does not perform probability redistribution. In the simplified case of uniform edge distribution on the vertices with mean probability $\tilde{p}$, MAE($\delta_A(u)$)$=\frac{\tilde{p}(1-\alpha)}{|V|}|E|$. Since all other factors are constant, MAE($\delta_A(u)$) increases linearly with $|E|$. \NI\ that applies limited probability redistribution yields smaller error. Clearly the winner again is \EMD\ with much smoother increase, verifying its robustness for dense graphs.   

\setcounter{figure}{9} 
\begin{figure*}[!hb]
    \begin{center}
        \centering
       \hspace*{-1cm}
        \subfigure[PR (Flickr)]{
            \includegraphics[width=0.26\linewidth]{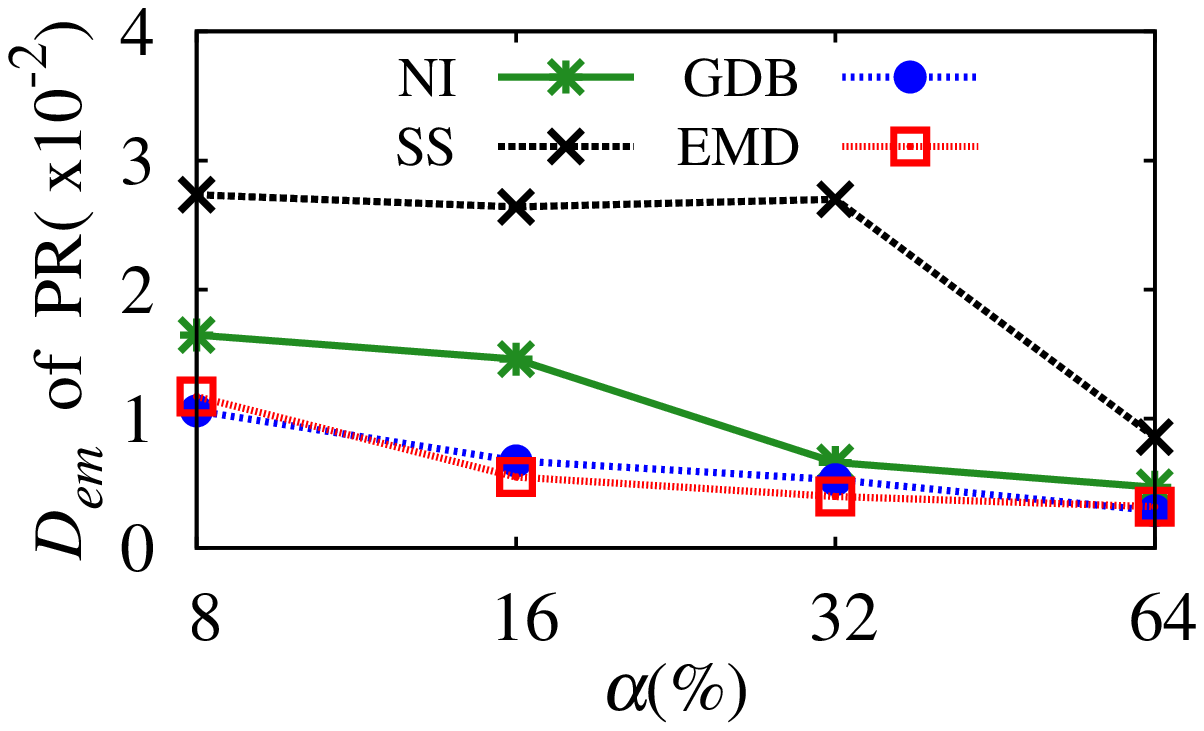}
        }
       \hspace*{-0.7cm}
        \subfigure[SP (Flickr)]{
            \includegraphics[width=0.26\linewidth]{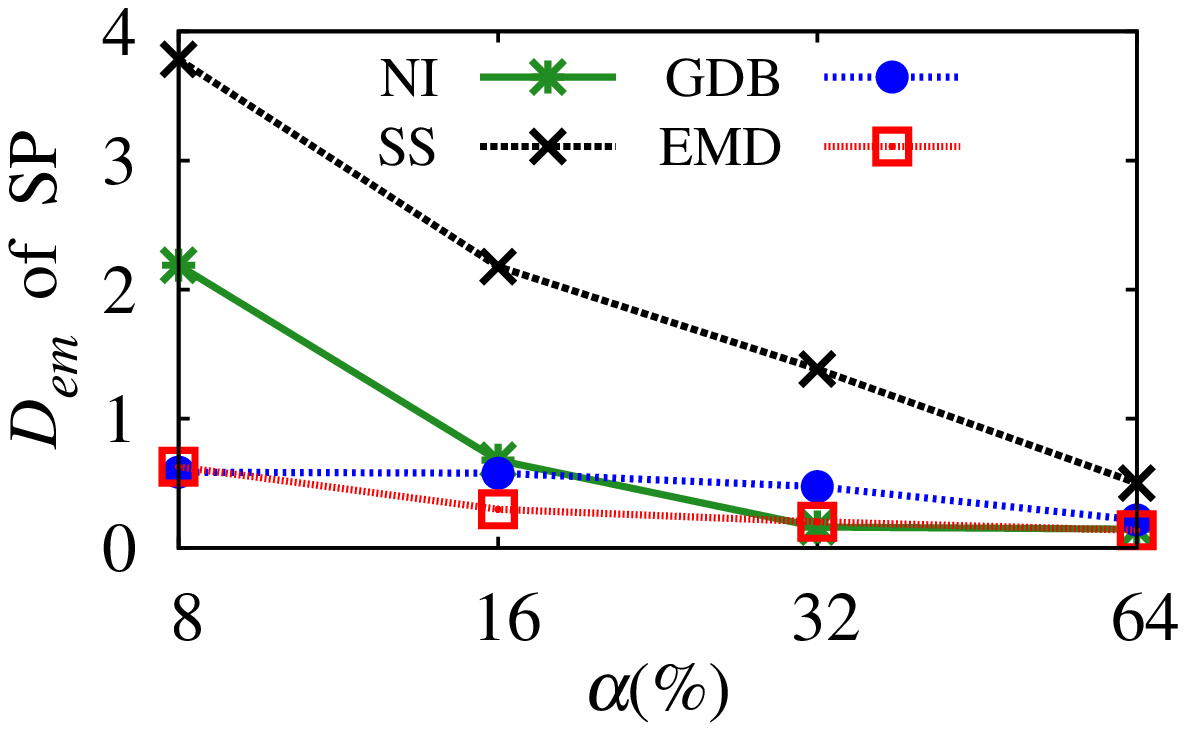}
        }
       \hspace*{-0.5cm}
         \subfigure[RL (Flickr)]{
            \includegraphics[width=0.26\linewidth]{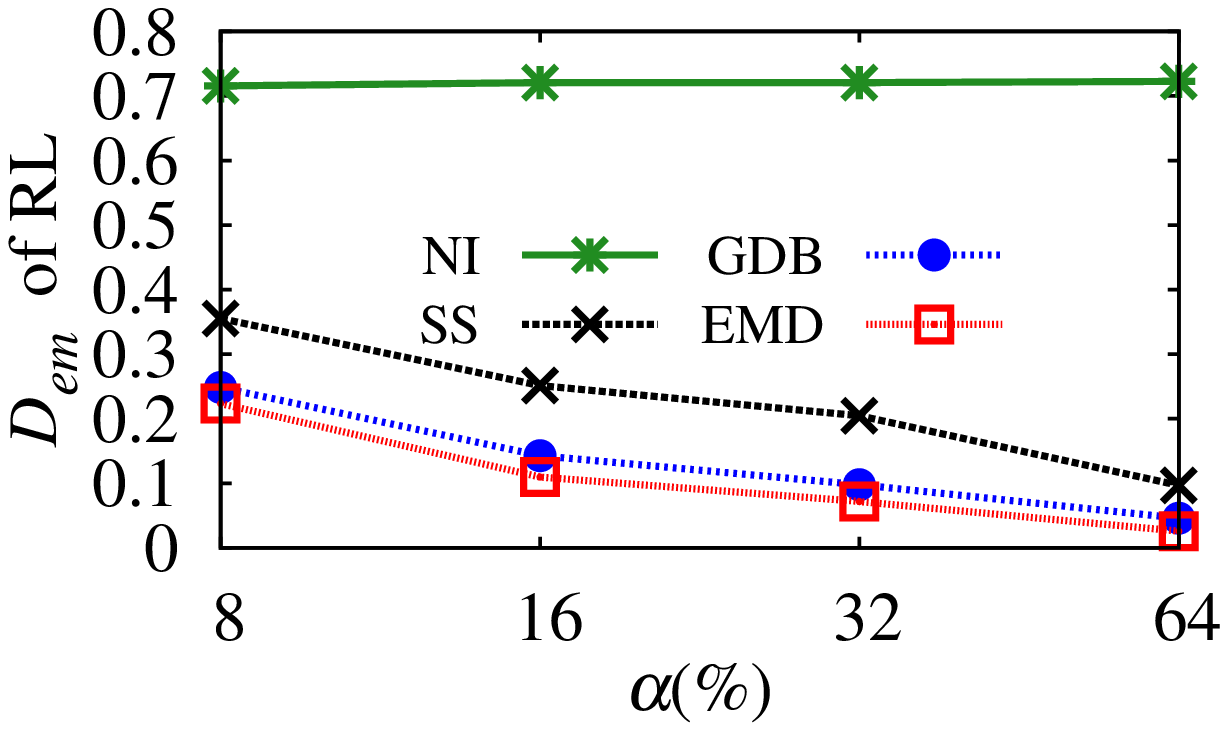}
        }
       \hspace*{-0.4cm}
         \subfigure[CC (Flickr)]{
            \includegraphics[width=0.26\linewidth]{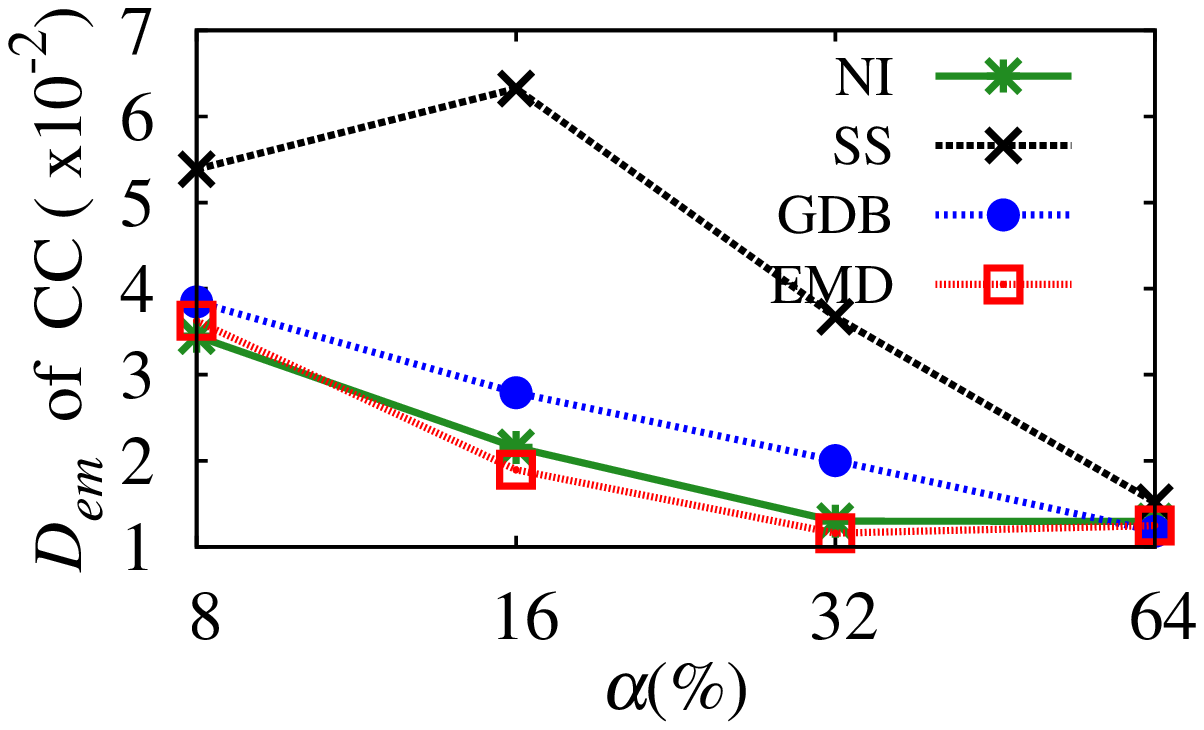}
        }

       \vspace{-4mm} 
       \hspace*{-1cm}
        \subfigure[PR (Twitter)]{
            \includegraphics[width=0.26\linewidth]{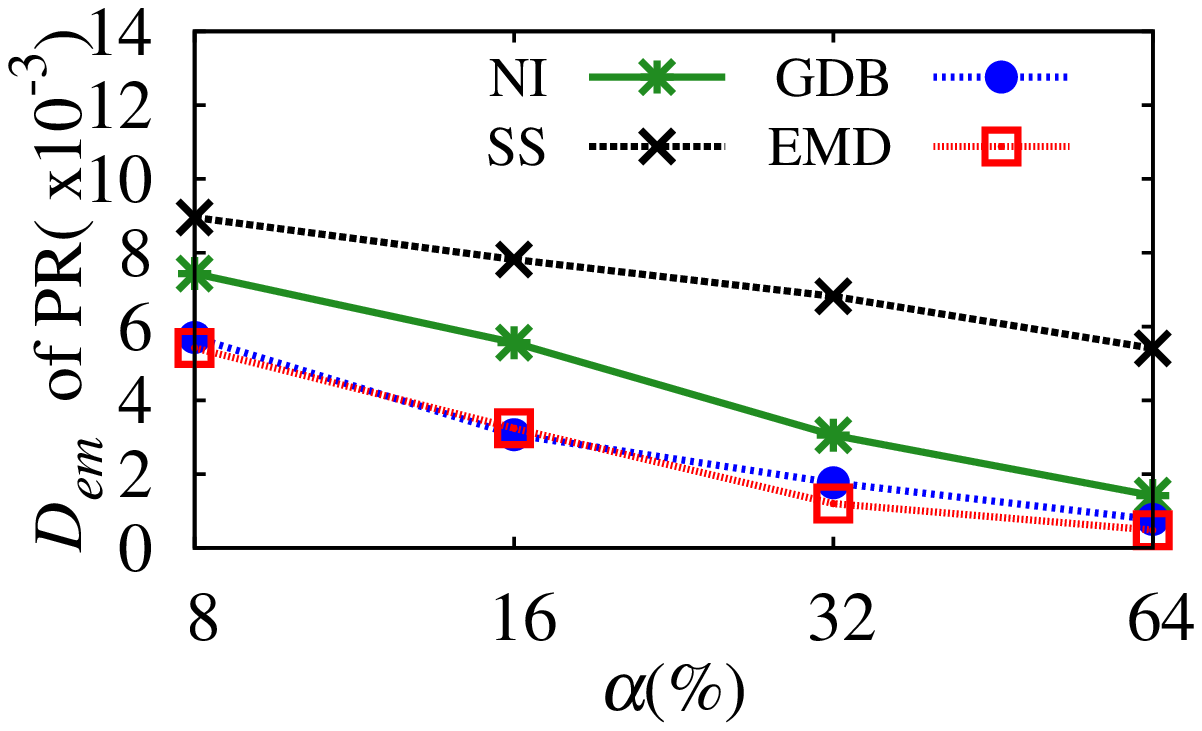}
        }\vspace{-1mm}
       \hspace*{-0.7cm}
        \subfigure[SP (Twitter)]{
            \includegraphics[width=0.26\linewidth]{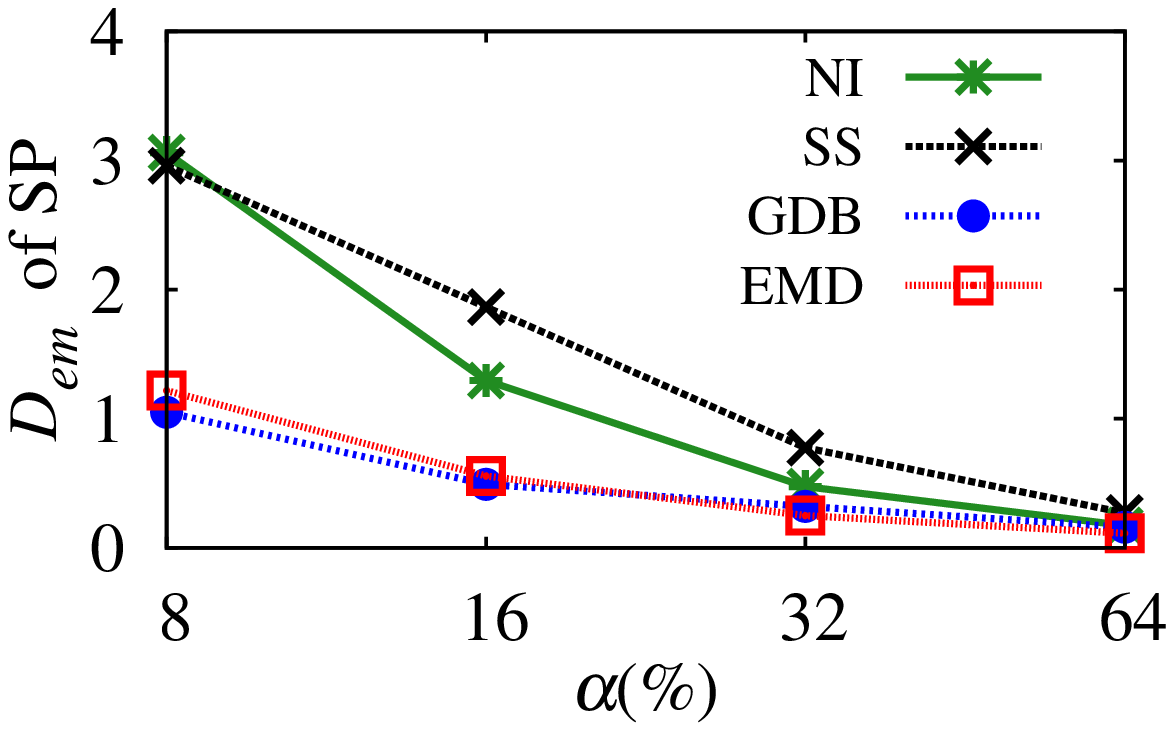}
        }\vspace{-1mm}
       \hspace*{-0.5cm}
         \subfigure[RL (Twitter)]{
            \includegraphics[width=0.26\linewidth]{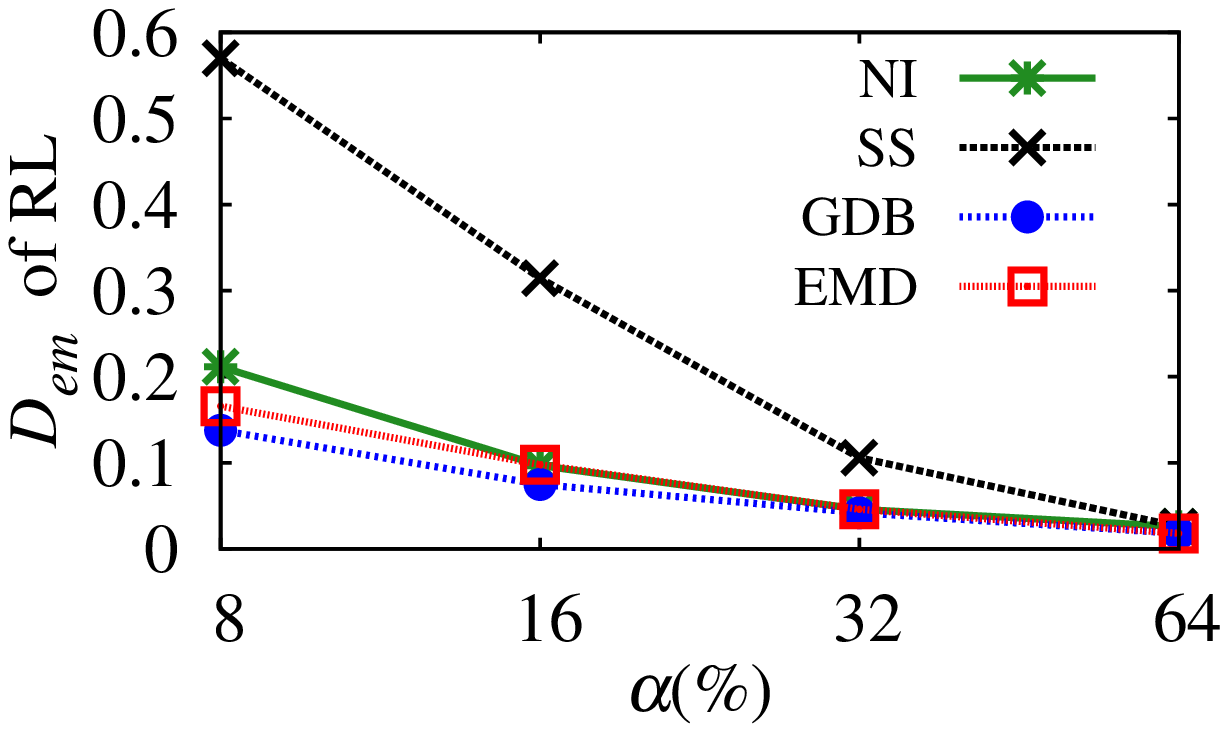}
        }\vspace{-1mm}
       \hspace*{-0.4cm}
         \subfigure[CC (Twitter)]{
            \includegraphics[width=0.26\linewidth]{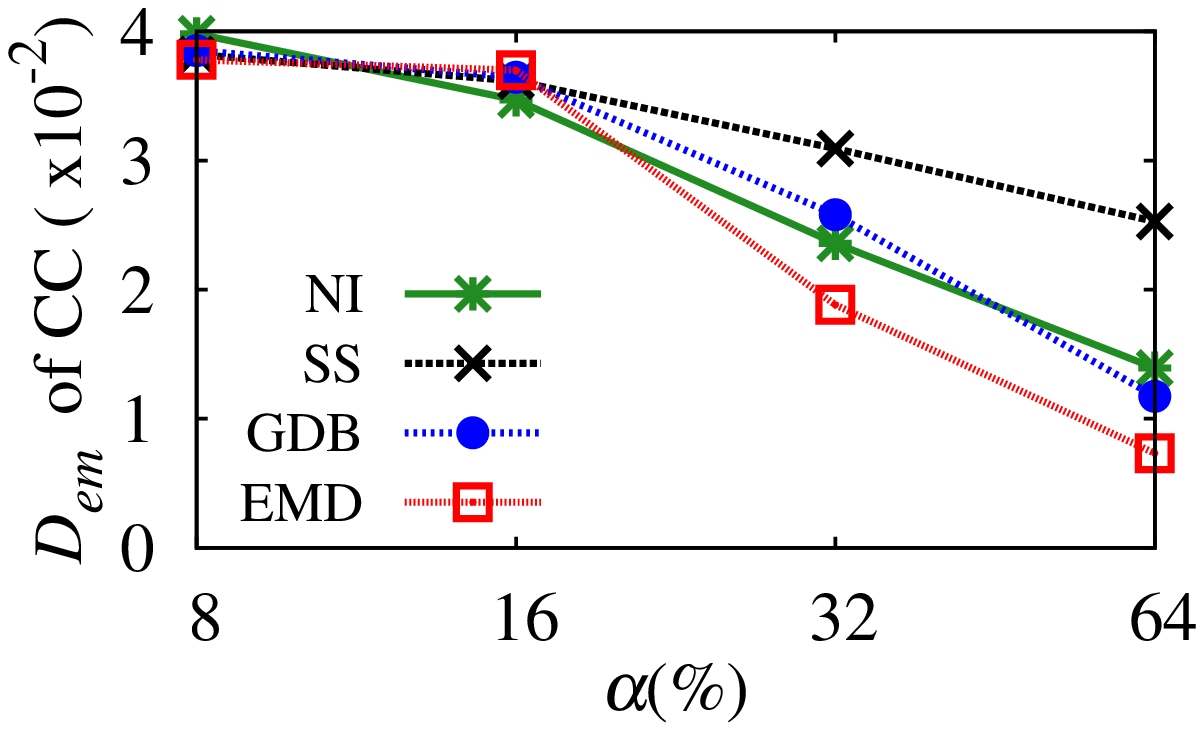}
        }
        \caption{Earth mover's distance $D_{em}$ for Pagerank (PR), Shortest Path distance (SP), Reliability (RL) and Clustering Coefficient (CC) (real datasets)}
        \label{plot:queries}
 \vspace{-0.5cm}
    \end{center}
\end{figure*}

Figure \ref{plot:entropy} plots the relative entropy of the sparsified graphs versus the sparsification ratio (real datasets) and the density (synthetic datasets). The relative entropy of a sparsified graph $\mathcal{G'}$ is the ratio $\frac{H(\mathcal{G'})}{H(\mathcal{G})}$, where $\mathcal{G}$ is the original graph. \EMD\ and \GDB\ have at least an order of magnitude less entropy for small $\alpha$ compared to \NI\ and \SS\, which overall perform similarly. This is expected since our methods aim at reducing entropy, unlike the competitors that are designed for deterministic graphs (zero entropy). Relative entropy increases with $\alpha$, always remaining less than 1.  Figure \ref{plot:entropy}(c) plots the entropy of the synthetic graphs with $\alpha=16\%$. The relative entropy is constant because the percentage of edges in the sparsified graph remains the same.
\setcounter{figure}{7} 
\begin{figure}[h!]
    \begin{center}
        \centering
        
        \hspace*{-0.45cm}
        \subfigure[Flickr]{
            \includegraphics[width=0.33\linewidth]{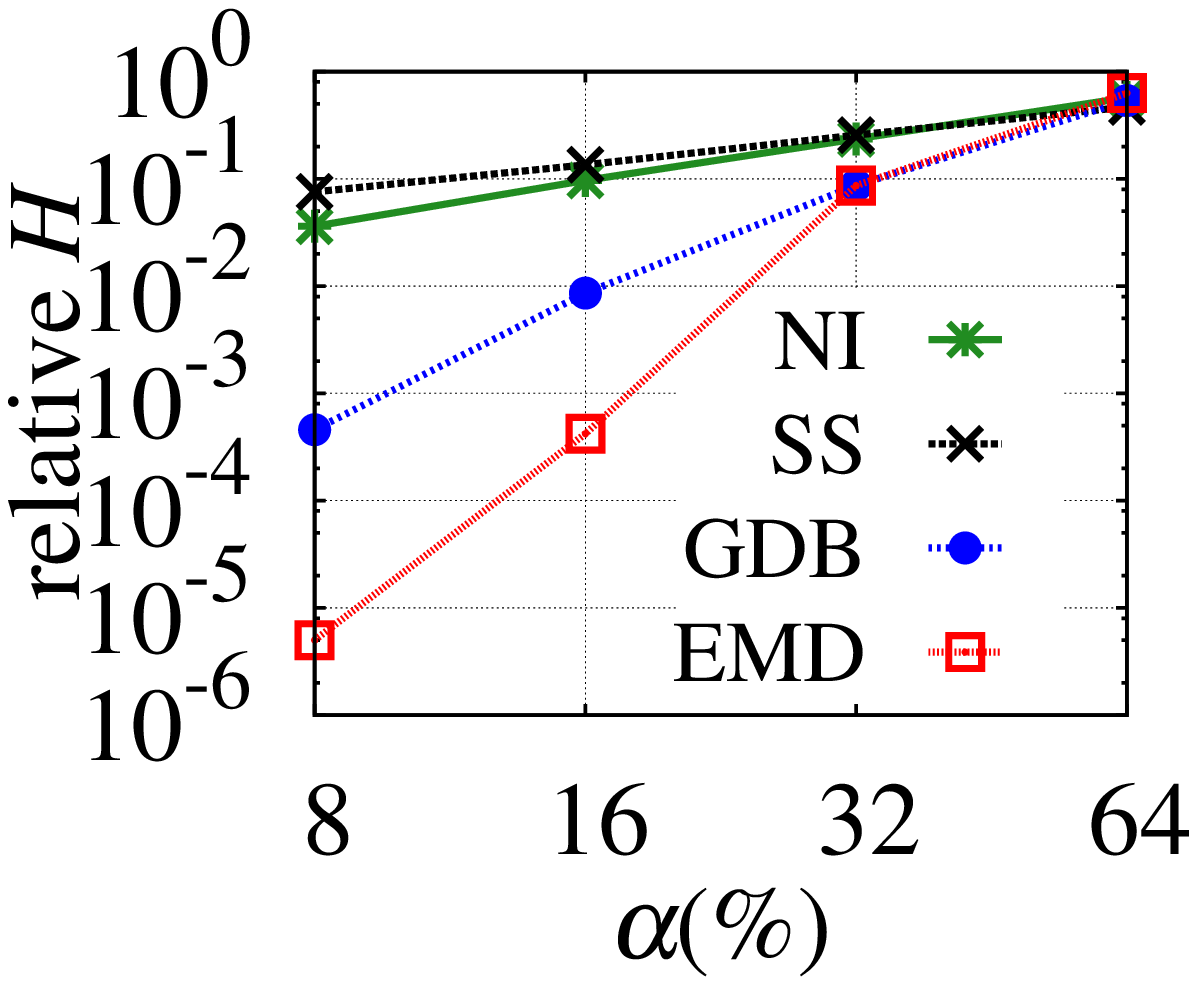}
        }\vspace{-2mm}
       \hspace*{-0.45cm}
        \subfigure[Twitter]{
            \includegraphics[width=0.33\linewidth]{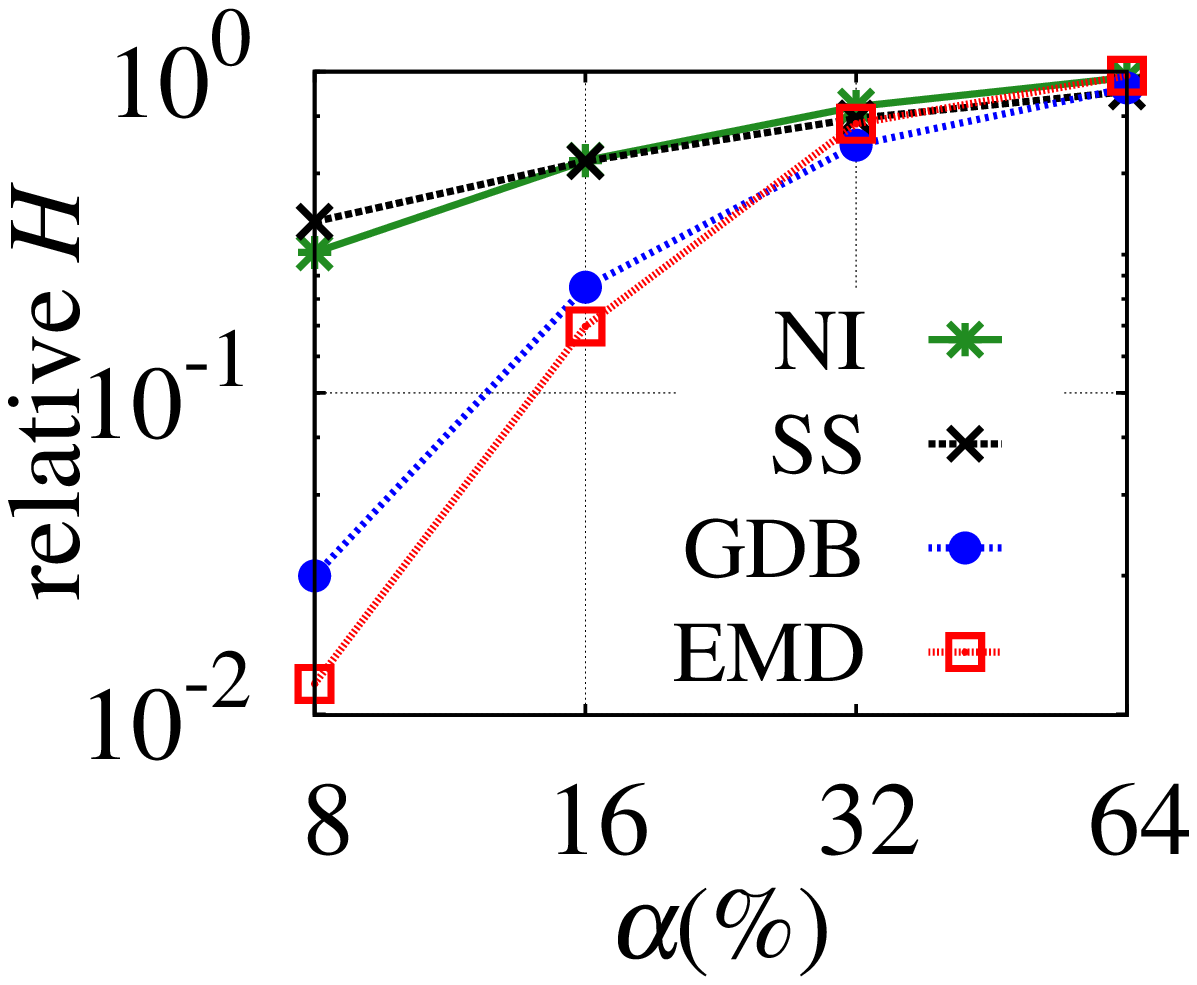}
        }
        \vspace{-1mm}
       \hspace*{-0.45cm}
        \subfigure[Synthetic]{
            \includegraphics[width=0.33\linewidth]{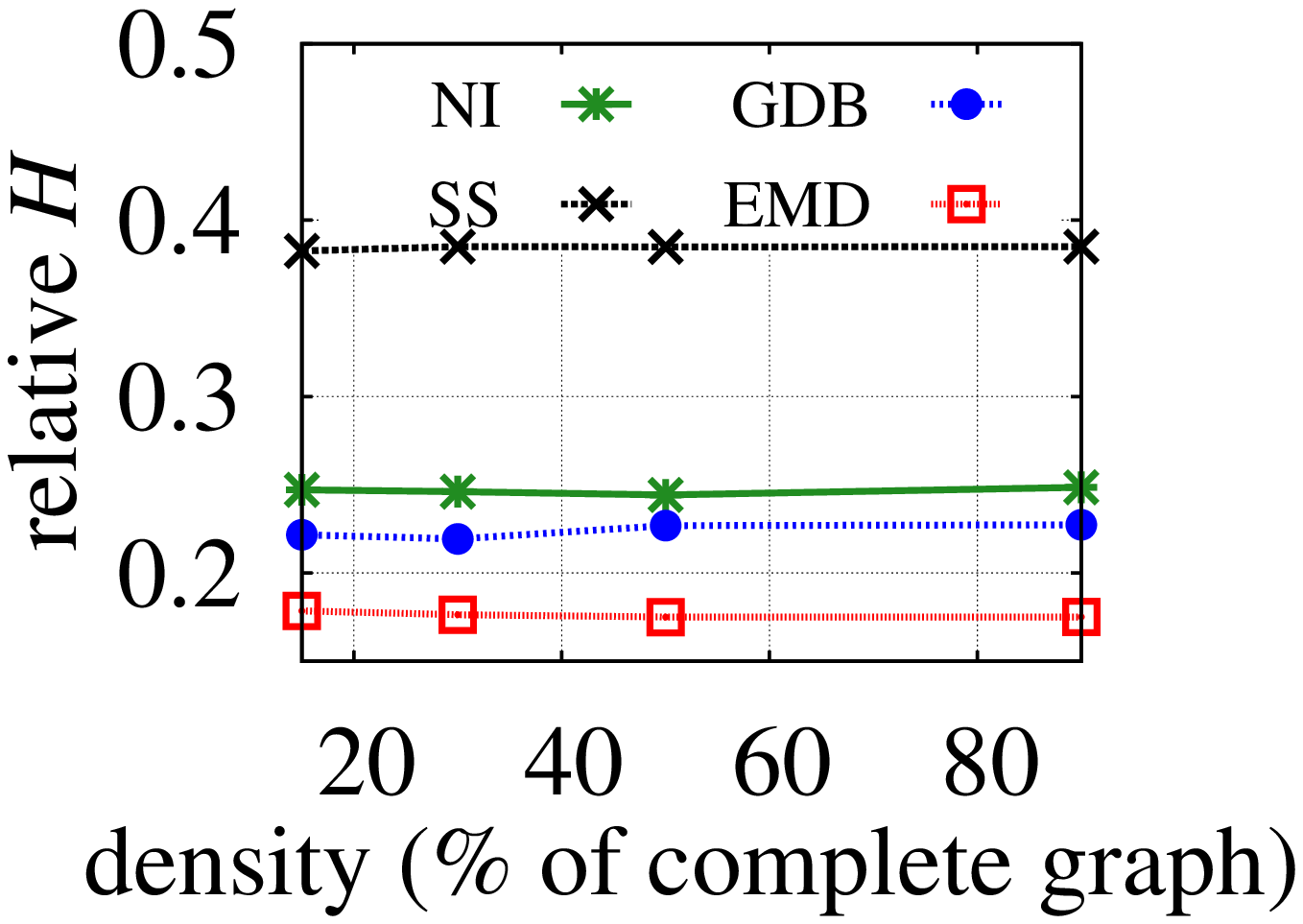}
        }
                
        \caption{Graph entropy $H$ (real and synthetic datasets)}
        \label{plot:entropy}
 \vspace{-0.5cm}
    \end{center}
\end{figure}

\begin{figure}[!h]
    \begin{center}
        \centering
       \hspace*{-0.45cm}
        \subfigure[Flickr]{
            \includegraphics[width=0.49\linewidth]{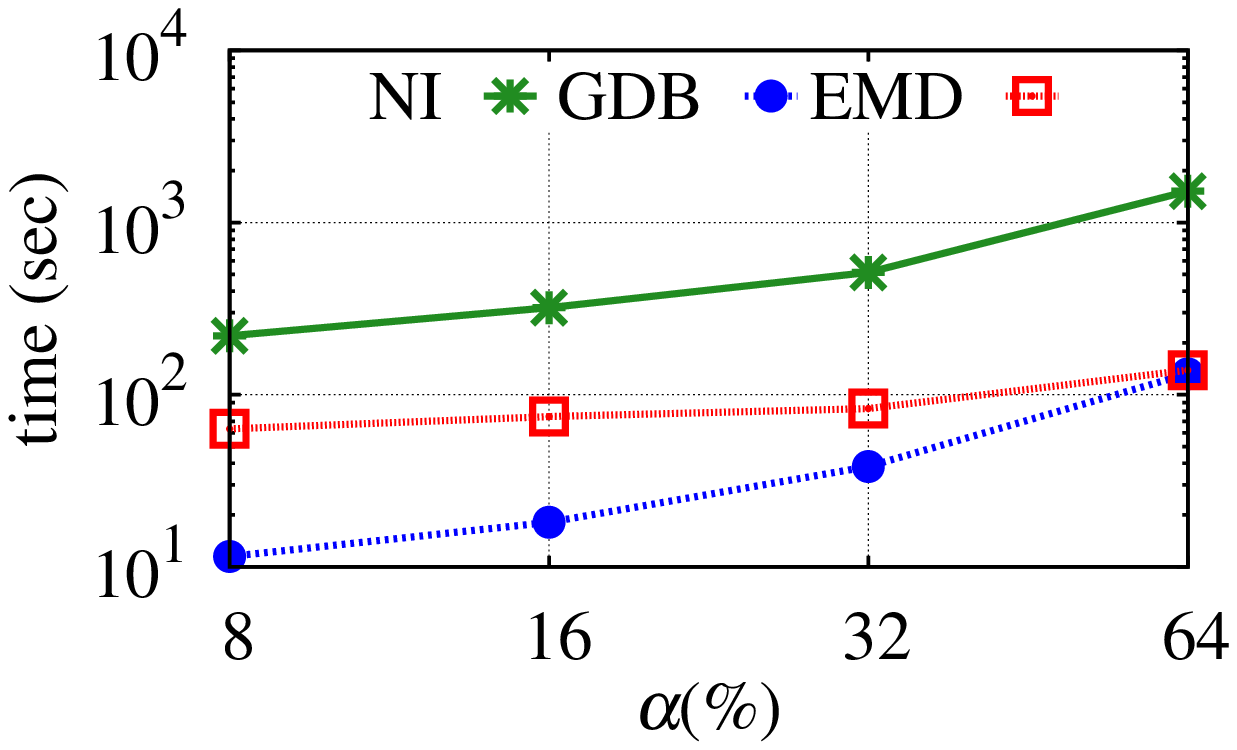}
        }
       \hspace*{-0.40cm}
        \subfigure[Twitter]{
            \includegraphics[width=0.49\linewidth]{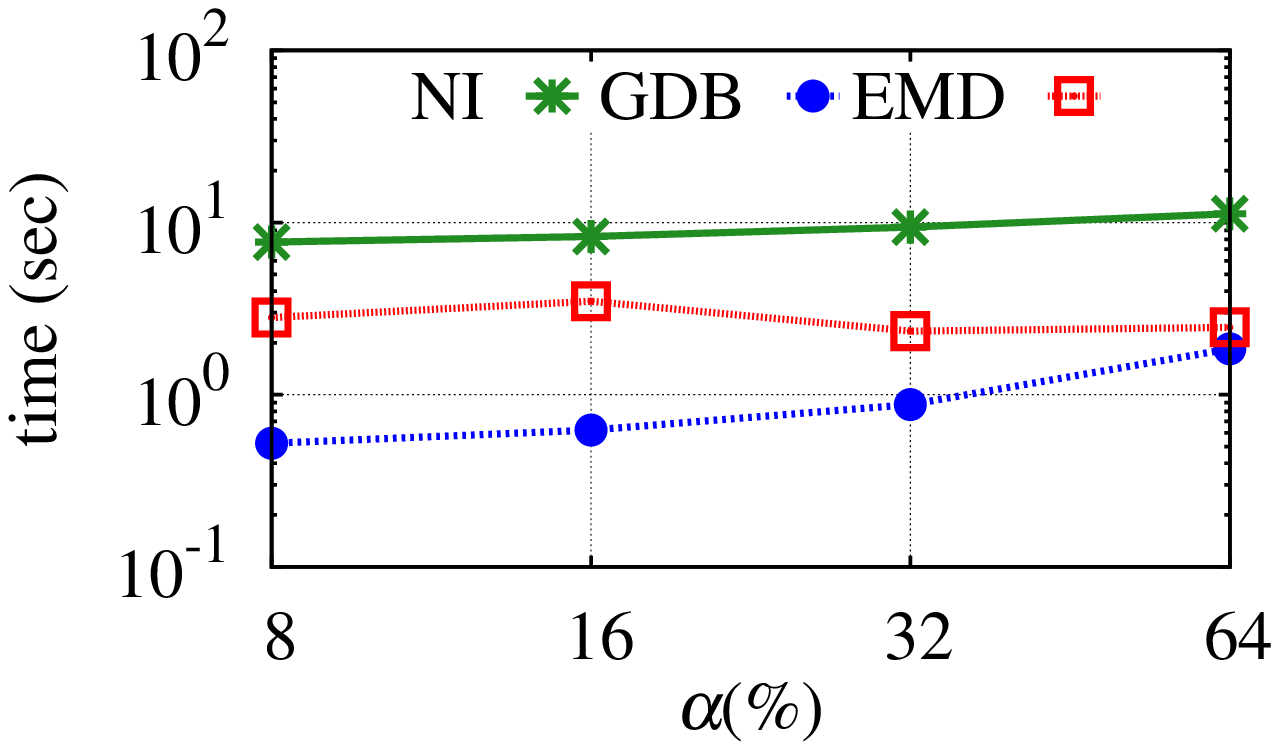}
        }
       \caption{Execution time (real graphs)}
        \label{plot:run}
 \vspace{-0.8cm}
    \end{center}
\end{figure}

The last experiment measures the running time of \GDB, \EMD\ and \NI\ in the real graphs. As shown in Figure \ref{plot:run}, the proposed methods usually terminate within a minute, whereas \NI\ is more than an order of magnitude slower. \SS\ is omitted from the diagrams because it requires several hours to terminate. Our methods scale linearly to the number of edges $\alpha|E|$ in the sparsified graph. In both plots of Figure \ref{plot:run}, we observe a linear increase with sparsification ratio $\alpha$. In addition, by combining the two plots we observe that for the same sparsification ratio $\alpha$, the running times of Flickr and Twitter differ by an order of magnitude, which is in accordance to the ratio of their number of edges $E$.

\subsection{Comparison with benchmarks on queries}\label{sec:expQueriesBench} 
\setcounter{figure}{11} 
\begin{figure*}[!bp]
    \begin{center}
        \centering
       \hspace*{-1cm}
        \subfigure[PR (Flickr)]{
            \includegraphics[width=0.26\linewidth]{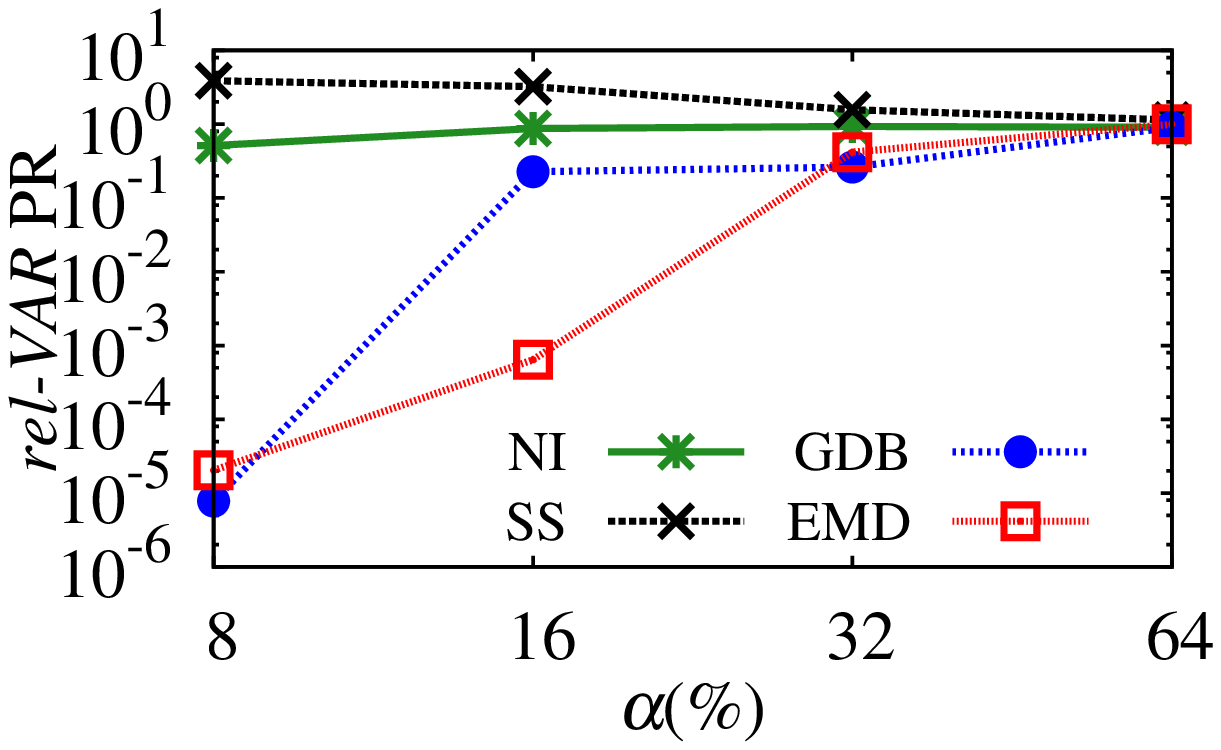}
        }
       \hspace*{-0.55cm}
        \subfigure[SP (Flickr)]{
            \includegraphics[width=0.26\linewidth]{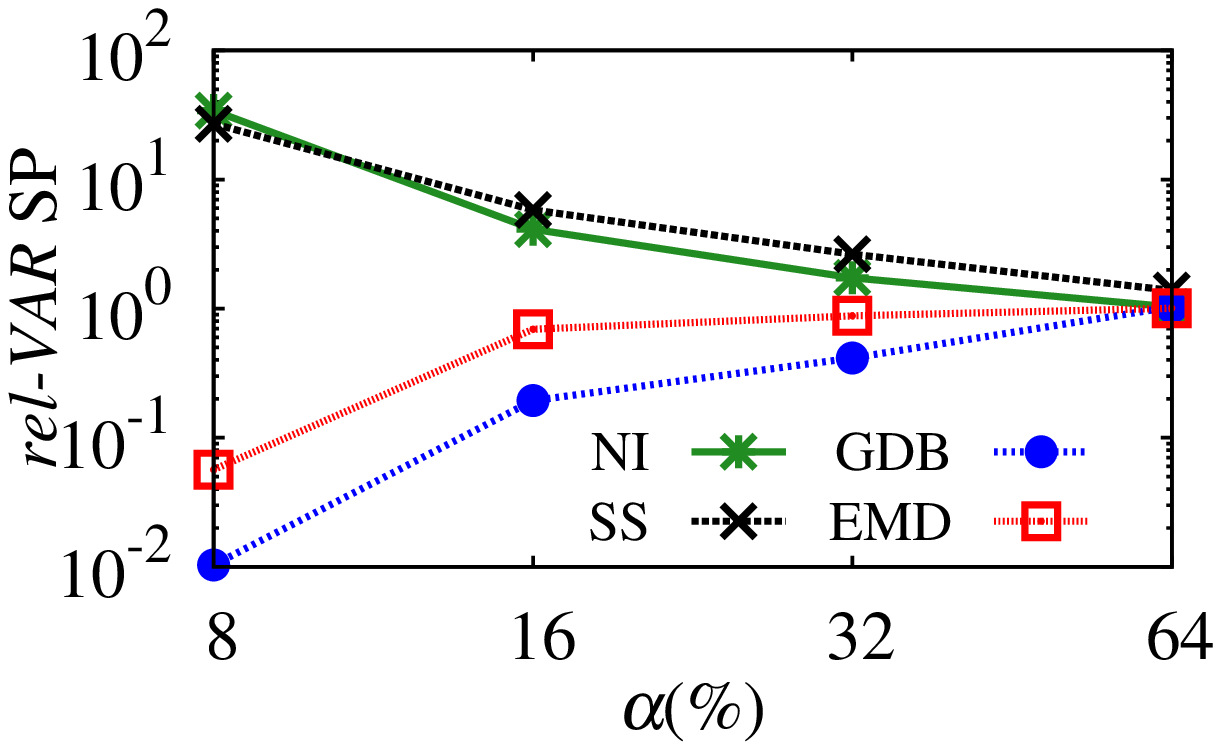}
        }
       \hspace*{-0.55cm}
         \subfigure[RL (Flickr)]{
            \includegraphics[width=0.26\linewidth]{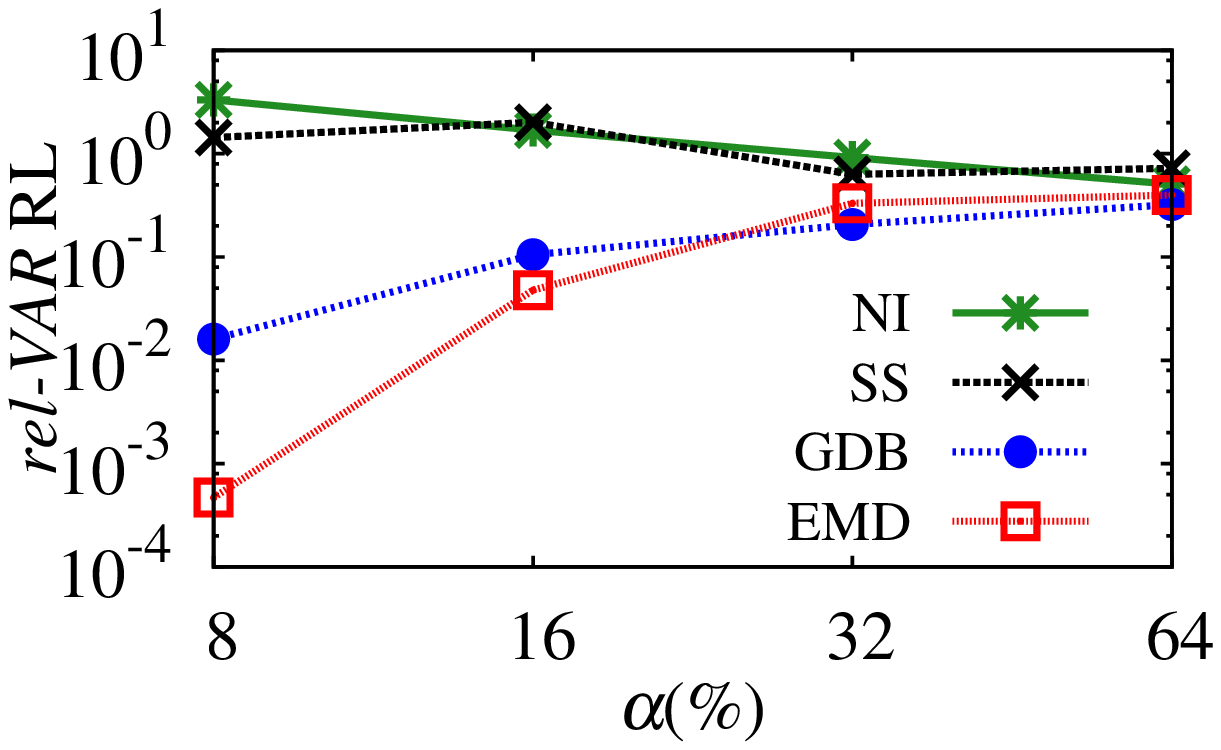}
        }
       \hspace*{-0.5cm}
         \subfigure[CC (Flickr)]{
            \includegraphics[width=0.26\linewidth]{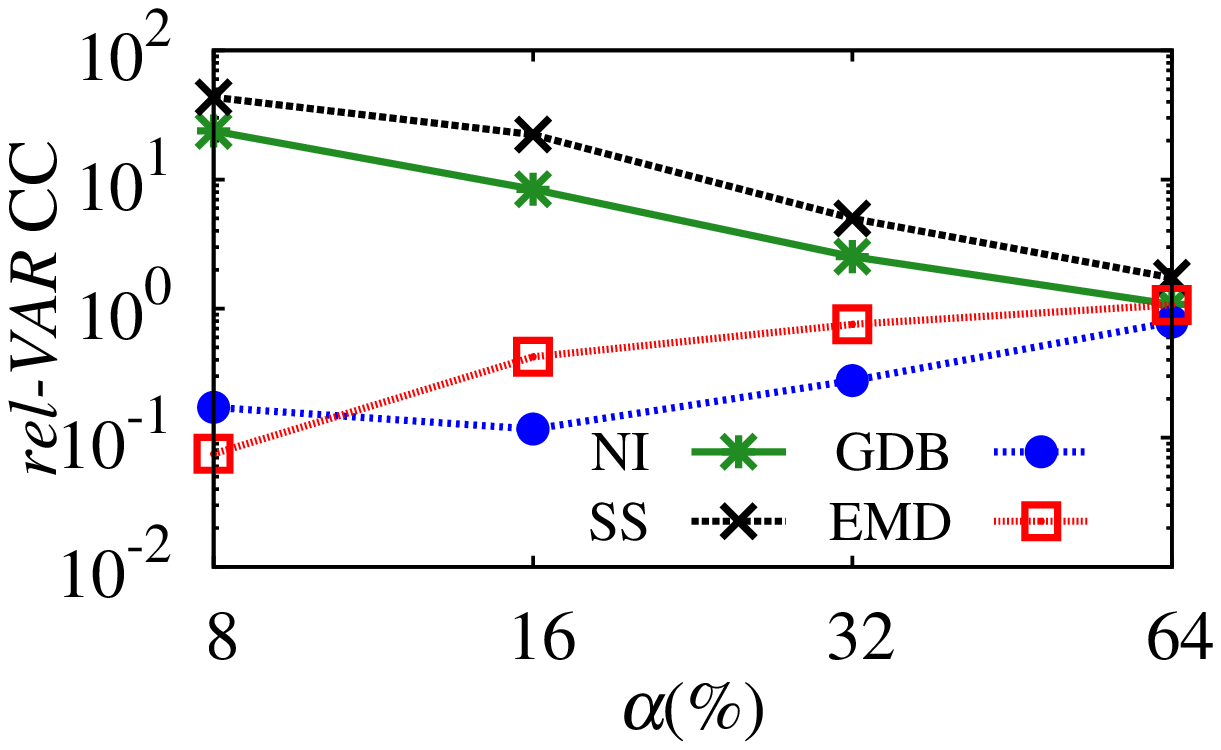}
        }

       \vspace{-4mm} 
       \hspace*{-1cm}
        \subfigure[PR (Twitter)]{
            \includegraphics[width=0.26\linewidth]{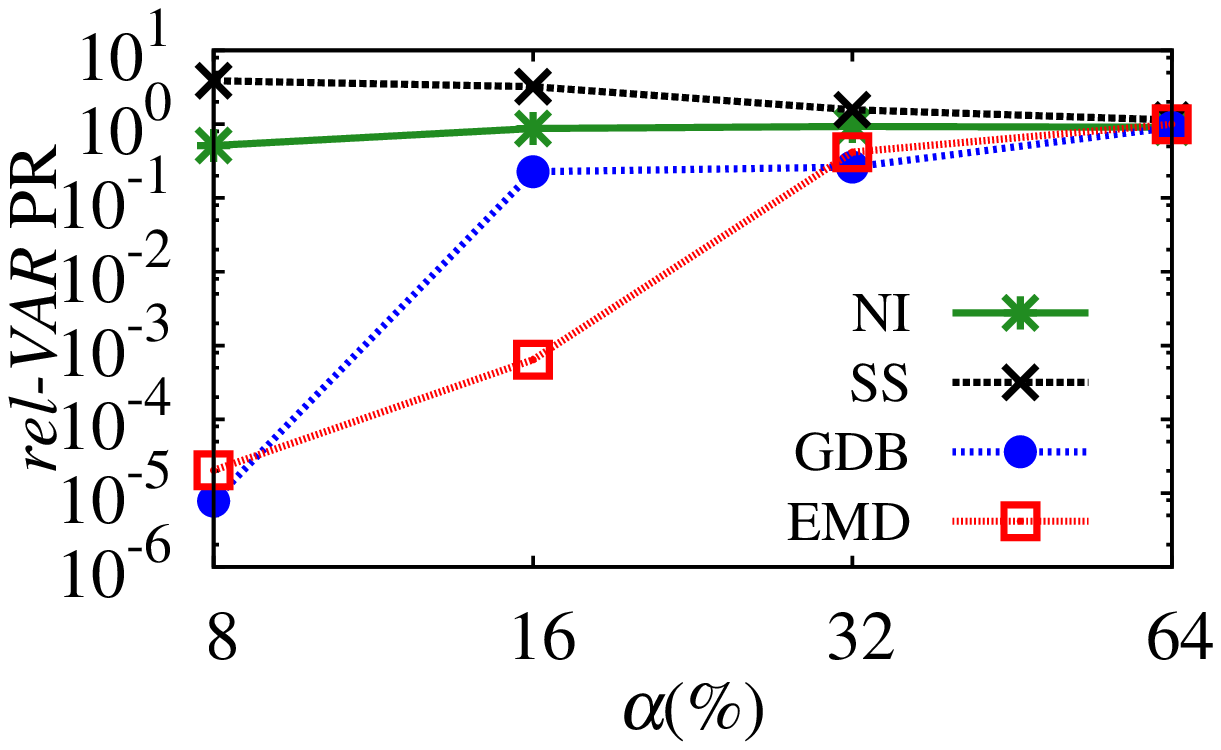}
        }\vspace{-1mm}
       \hspace*{-0.6cm}
        \subfigure[SP (Twitter)]{
            \includegraphics[width=0.26\linewidth]{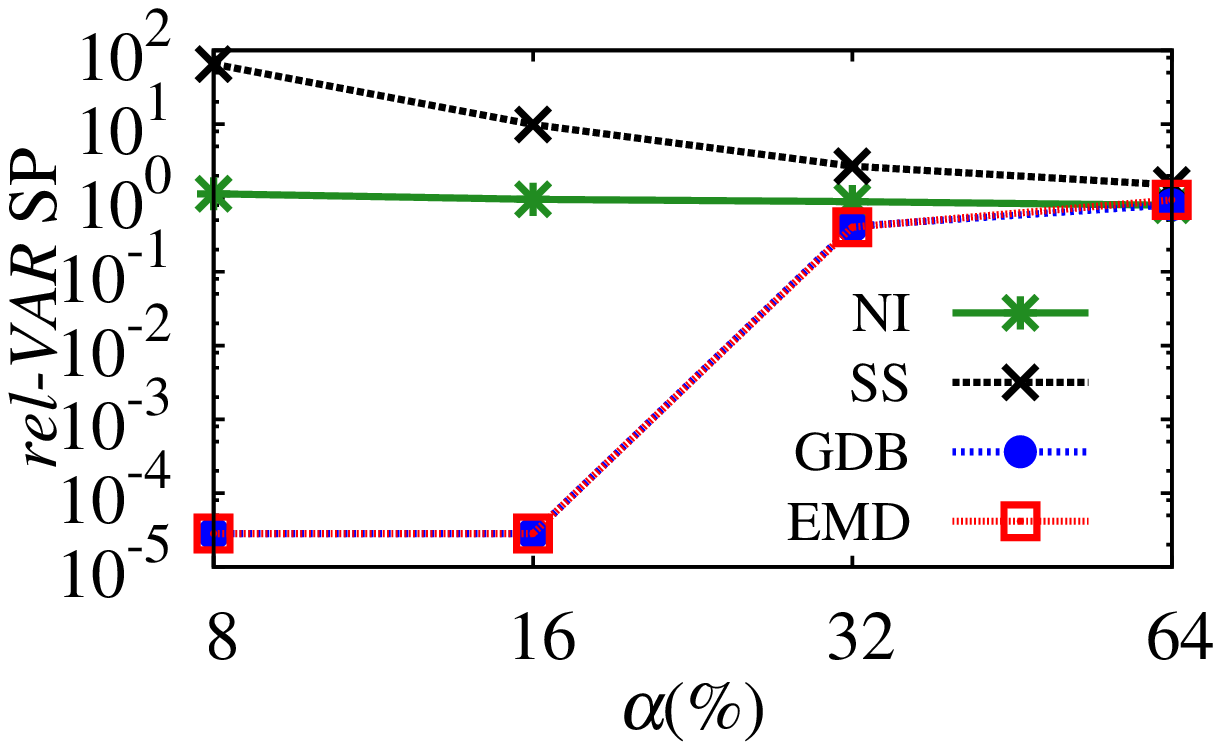}
        }\vspace{-1mm}
       \hspace*{-0.45cm}
         \subfigure[RL (Twitter)]{
            \includegraphics[width=0.26\linewidth]{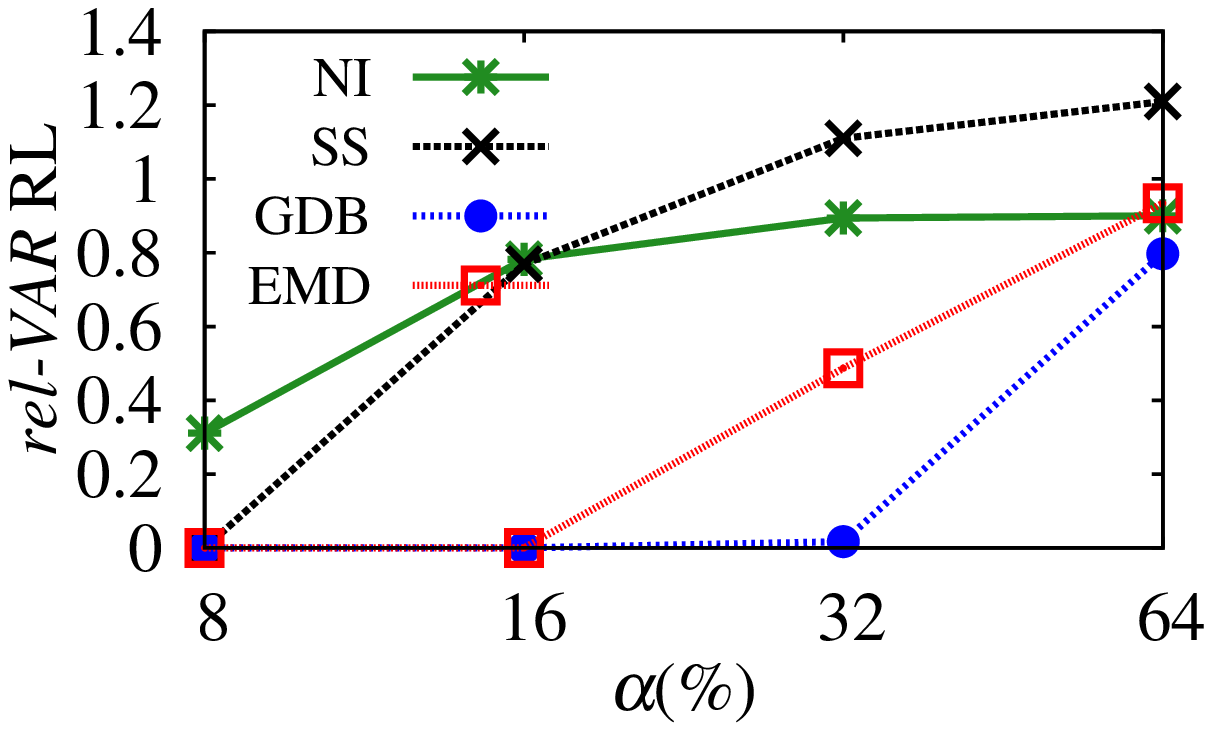}
        }\vspace{-1mm}
       \hspace*{-0.55cm}
         \subfigure[CC (Twitter)]{
            \includegraphics[width=0.26\linewidth]{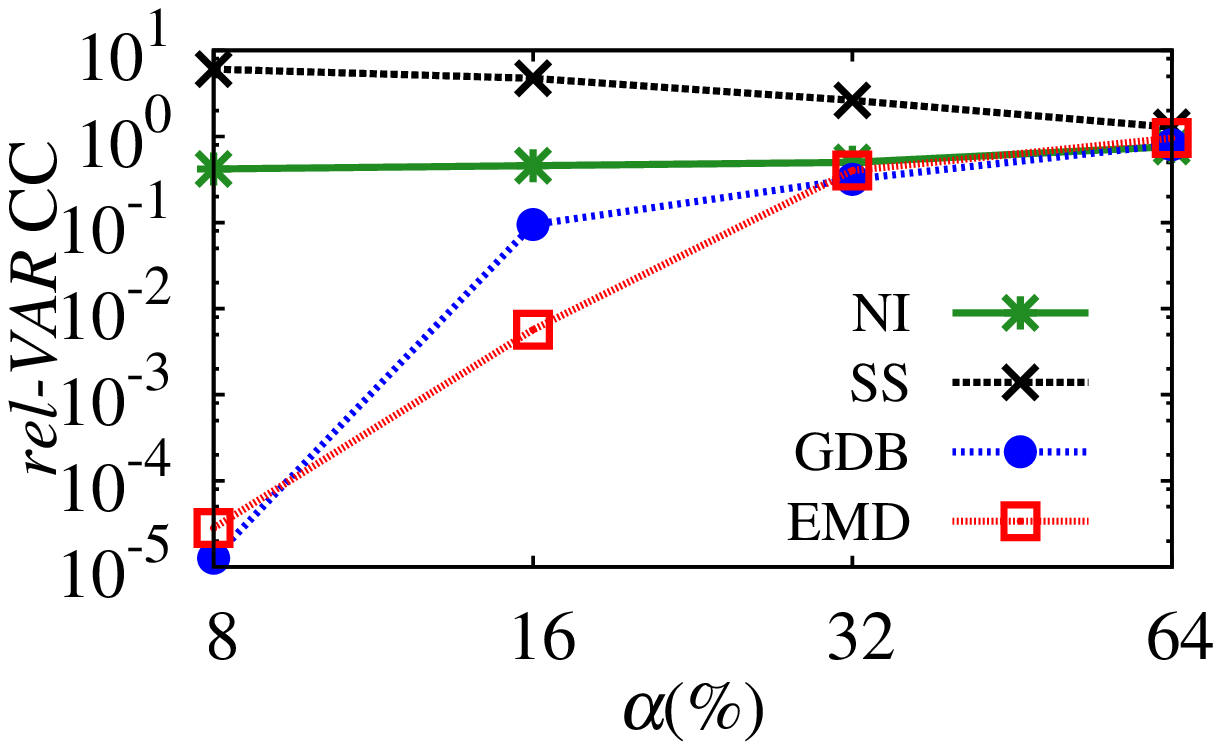}
        }
%
        \caption{Relative variance for Pagerank (PR), Shortest Path distance (SP), Reliability (RL) and Clustering Coefficient (CC)}
        \label{plot:queries_var}
 \vspace{-0.5cm}
    \end{center}
\end{figure*}  

We evaluate the following graph queries. (i) \textit{Pagerank} (PR) is a measure of the node's influence in the graph and has been widely used to rank Web page search results according to their links \cite{pagerank}. (ii) \textit{Shortest path distance} (SP) is the average shortest distance between a pair of vertices in all worlds excluding the ones that disconnect them. SP is essential for any task involving shortest path computations \cite{PBGK10}. (iii) \textit{Reliability} (RL) is the probability that a vertex is reachable from another in the graph. It is a common metric for the resilience of router networks. (iv) \textit{Clustering coefficient} (CC) is the ratio of the number of edges between the neighbours of a vertex to the maximum number of such links. CC constitutes an important metric for search strategies and social networks~\cite{Kossinets06012006}. Although we evaluate CC and PR on \textit{all} vertices of $\mathcal{G}$, we choose 1000 random vertex pairs for the evaluation of SP and RL, because the evaluation on all pairs would be too expensive to terminate for our datasets.

\smallskip \noindent \textbf{Quality results.}
Let $\mathcal{G'}$ be a sparsified subgraph of $\mathcal{G}$, and a query $Q$. $Q$ is evaluated through Monte-Carlo sampling on 500 possible worlds of both $\mathcal{G'}$ and $\mathcal{G}$. The various outcomes of $Q$ in different samples of $\mathcal{G'}$ form a cumulative distribution $F_{\mathcal{G'},Q}(x)$ of results. To quantify the similarity of $\mathcal{G'}$ to $\mathcal{G}$ with respect to $Q$, we have to measure the difference between $F_{\mathcal{G'},Q}(x)$ and $F_{\mathcal{G},Q}(x)$. To this end, a robust metric is the \textit{earth mover's distance} $D_{em}$ \cite{earth}. Intuitively, $D_{em}$ measures the minimum change that aligns the two distributions. Formally, let $\{x_0,x_1,\cdots,x_M\}$ be the ordered set of all observed results of $Q$ in $\mathcal{G}$ and $\mathcal{G'}$. To compute $D_{em}$ we apply the following equation:
\begin{equation}\label{earthdist}
	D_{em}(\mathcal{G}, \mathcal{G'}, Q) = \sum_{i=1}^{M}\big| F_{\mathcal{G},Q}(x_i)-F_{\mathcal{G'},Q}(x_i)\big|\cdot (x_i- x_{i-1})
\end{equation} 
\vspace{-0.1in}

Figure \ref{plot:queries} plots $D_{em}$ versus the sparsification ratio. Each row of diagrams corresponds to a dataset and each column to a query. With few exceptions, \GDB\ and \EMD\ outperform the benchmarks for all settings, usually by a wide margin. Moreover, the diagrams are consistent with those on structural properties, confirming the correlation of our objective functions with the performance of the sparsified graphs for diverse queries. \SS\ yields the highest error even on the SP metric, which constitutes its focus. The main cause for its poor performance is that it does not involve any probability redistribution. Although \NI\ achieves good approximation for CC, it usually introduces large error for the rest of the queries. \EMD\ is the winner for high sparsification ratio, while \GDB\ is preferable for small $\alpha$ in most queries. This is in accordance with Figure \ref{plot:objective}, where \GDB\ preserves better the structural properties for $\alpha=8\%$. 

Figure \ref{plot:queries_dense}(a) (resp. Figure \ref{plot:queries_dense}(b)) illustrates $D_{em}$ of PR (resp. SP) on the synthetic datasets, as a function of density for $\alpha=16\%$. The proposed techniques clearly yield smaller error than the benchmarks. Observe that PR is node centric and highly correlated with the degree; thus, the diagram of Figure \ref{plot:queries_dense}(a) is similar to that of Figure \ref{plot:objective_dense}(a). The plots of CC are similar to those of PR and omitted.
On the other hand, the error of SP decreases with increasing density because more alternative short paths are available, due to the abundance of edges. RL has practically zero error for all methods, since the dense graph has reliability almost 1 for all pairs of vertices. 

\setcounter{figure}{10} 
 \vspace{-0.2cm}
\begin{figure}[h!]
    \begin{center}
        \centering
        
        \hspace*{-0.45cm}
        \subfigure[PR (Synthetic)]{
            \includegraphics[width=0.51\linewidth]{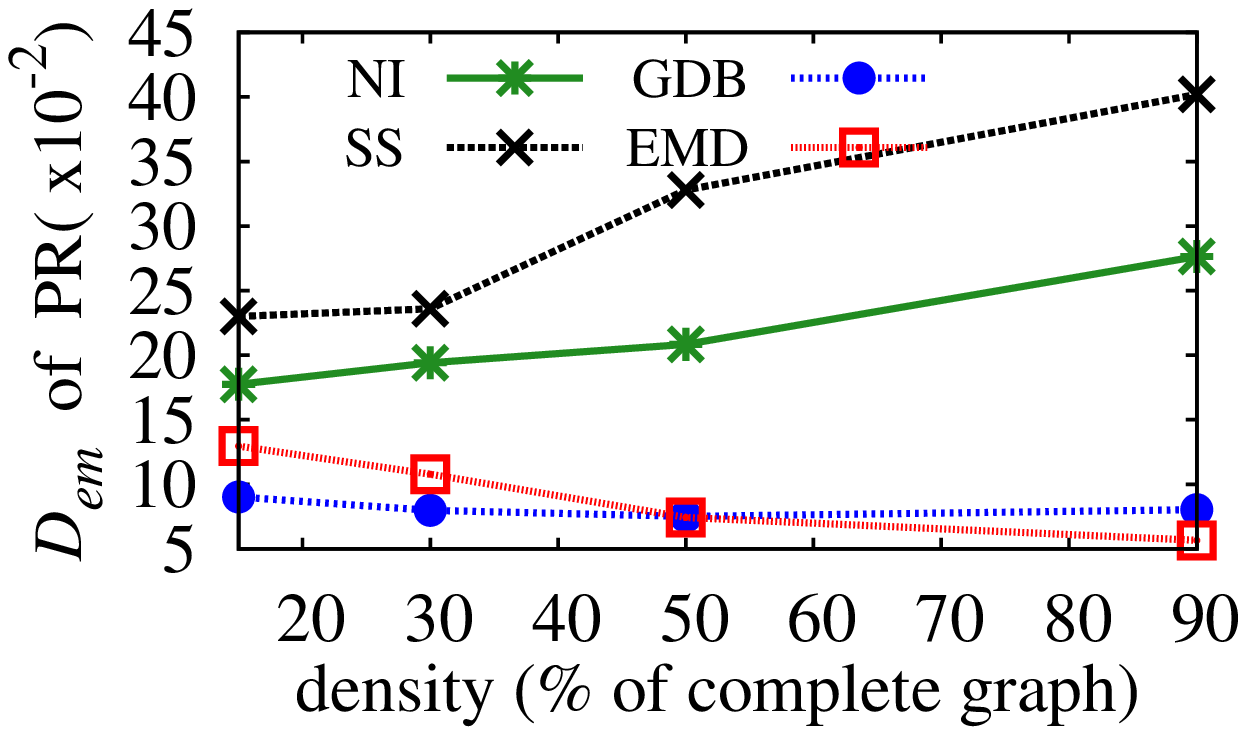}
        }
       \hspace*{-0.45cm}
        \subfigure[SP (Synthetic)]{
            \includegraphics[width=0.51\linewidth]{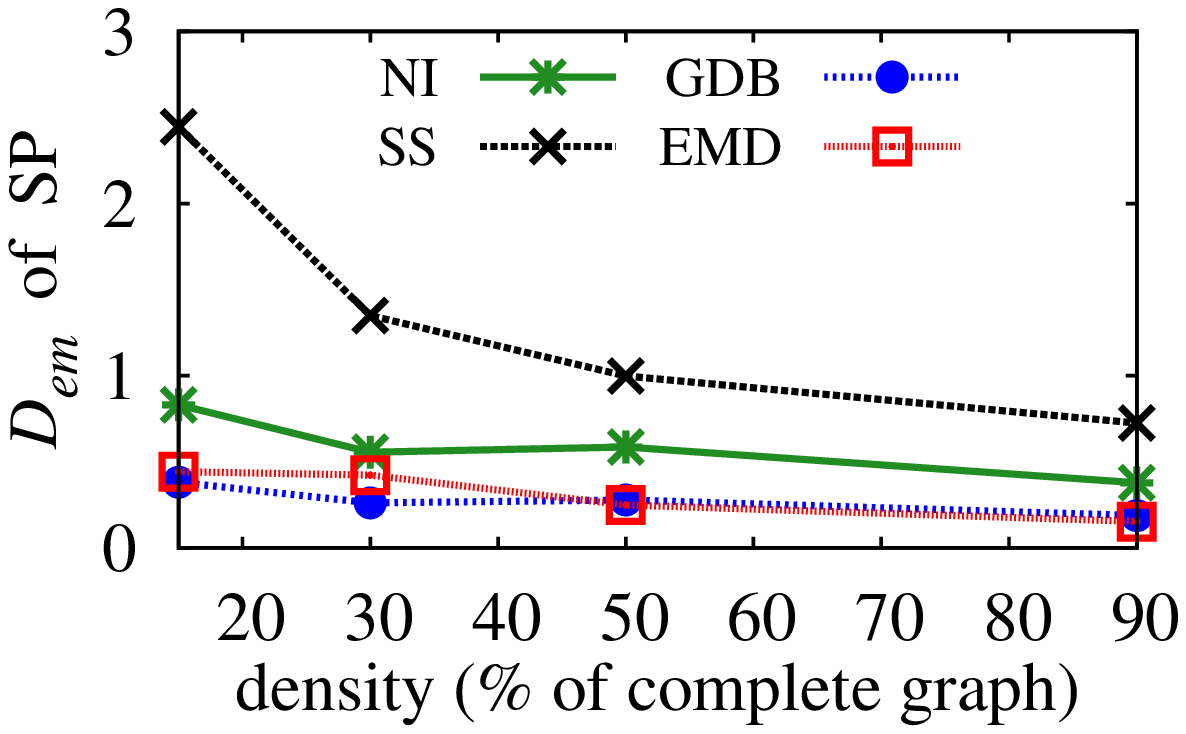}
        }
                
       \caption{Earth mover's distance $D_{em}$ for Pagerank (PR) and Shortest Path distance (SP) (synthetic datasets)}
        \label{plot:queries_dense}
\vspace{-0.3cm}
    \end{center}
\end{figure}

\smallskip \noindent \textbf{Variance results.}
We assess the performance of the various sparsifiers with respect to the variance of an MC estimator on the above queries. Specifically, due to the randomized nature of MC estimators, different executions of the same experiment may yield different results. The variance quantifies the deviation of results from the mean. Let $\Phi(\mathcal{G})$ denote the result of an MC simulation on an uncertain graph $\mathcal{G}$. 

A sparsified graph with low variance on MC estimator\footnote{Slightly abusing notation, we use the term "variance of $\mathcal{G}$" to imply the variance of a query estimator on a graph $\mathcal{G}$.} implies the need of fewer samples for accurate estimation. Specifically, according to the theory of MC simulations, the unknown expected value of a query in $\mathcal{G}$ belongs to the \textit{confidence interval} $CI= [\Phi(\mathcal{G})-1.96\sigma(\mathcal{G})/\sqrt{N}, \Phi(\mathcal{G})+1.96\sigma(\mathcal{G})/\sqrt{N}]$ with probability 95\%, where $\sigma(\mathcal{G})$ is the variance of the query in $\mathcal{G}$ and $N$ is the number of samples. Let the \textit{confidence width} $CW$ be the length of the confidence interval, i.e., $CW=2CI=3.92\sigma(\mathcal{G})/\sqrt{N}$. In order to achieve the same level of accuracy $CW$ between the original graph $\mathcal{G}$ and the sparsified $\mathcal{G'}$, we require $CW=CW'\rightarrow \sigma(\mathcal{G})/\sqrt{N}=\sigma(\mathcal{G'})/\sqrt{N'} \rightarrow N'/N= \big( \sigma(\mathcal{G'})/\sigma(\mathcal{G})\big)^2$. Intuitively, small relative variance $\frac{\sigma(\mathcal{G'})}{\sigma(\mathcal{G})}$  implies the need of fewer samples for accurate estimation.  This is why variance is among the most important metrics for the quality of MC simulation (see \cite{varianceReduction,li2014ICDE}). However, 
calculating the actual variance is intractable. Thus, we follow a strategy similar to \cite{li2014ICDE} for an unbiased estimator of the variance of $\Phi(\mathcal{G})$, denoted as $\hat{\sigma}(\mathcal{G})$. Specifically, we run each estimator ($\Phi_i(\mathcal{G})$) 100 times. Then the unbiased estimator of the variance of $\Phi(\mathcal{G})$ is $\hat{\sigma}(\mathcal{G})=\sum_{i=1}^{100} (\Phi_i(\mathcal{G})-\bar{\Phi}(\mathcal{G}))^2/99$, where $\bar{\Phi}(\mathcal{G})$ denotes the mean of $\Phi_i(\mathcal{G})$ for $(i=1, \cdots ,100)$.


Figure \ref{plot:queries_var} illustrates the relative variance of the queries versus the sparsification ratio $\alpha$. Let $\hat{\sigma}(\mathcal{G})$ and $\hat{\sigma}(\mathcal{G'})$ denote the variance of the MC estimator on the original and the sparsified graph respectively. The y-axis represents the relative variance, i.e., the ratio $\frac{\hat{\sigma}(\mathcal{G'})}{\hat{\sigma}(\mathcal{G})}$. Each row of diagrams corresponds to a dataset and each column to a query. Consistently, \EMD\ and \GDB\ drop the variance of the original graph up to several orders of magnitude. On the other hand, \NI\ and \SS\ have, in most cases, higher variance than the original graph. The justification of this result is based on the fact that \NI\ and \SS\ perform limited (if any) probability redistribution. Assume for instance the SP query. During evaluation we only consider \textit{reliable} possible worlds for which the distance of the query points is less than infinite. Sparsification without probability redistribution reduces the number of possible worlds for which a path exists. Consequently, a bigger proportion of possible worlds is discarted, increasing the variance of the estimator, which now depends on fewer samples. Moreover, according to the small world phenomenon, the majority of shortest paths are small (i.e., less than 10). Thus, the removal of edges is more likely to affect the the probability of short paths. This in turn increases the probability of larger paths, inflating the variance.

Our techniques alleviate the above short-comings by applying aggressive probability redistribution that preserves the expected number of edges, reducing the entropy of the sparsified graphs. This results in many edges having probability one. For instance, in Twitter with $\alpha=8\%$, 75\% of the edges of \GDB\ have probability 1. In comparison, in \NI\ only 25\% of the edges are deterministic. As $\alpha$ increases, fewer edges reach probability 1; thus, the variance of \EMD\ and \GDB\ increases. This result is very important and highlights one of the core differences of our methods compared to the competitors: the goal for entropy reduction. 

Summarizing the experiments, as shown in Table \ref{tab:ours} and Figure \ref{fig:ours}, the proposed techniques capture well the structural properties of the input uncertain graph, even if they do not constitute their explicit optimization criterion. For instance, variants that aim at the relative discrepancy $\delta_R$, e.g. $\EMD^R$-$t$, also preserve the absolute one $\delta_A$. According to Figures \ref{plot:queries}-\ref{plot:queries_dense}, the preservation of structural properties leads to accurate results for various queries with different characteristics. Moreover, reducing the entropy of the uncertain graph (Figure \ref{plot:entropy}), our methods decrease the variance of the MC estimator of all evaluated queries (Figure \ref{plot:queries_var}). This has huge effect in processing time, as considerably less samples are required for accurate query estimation.  As opposed to the proposed methods, techniques based on deterministic sparsification usually fail, both in terms of result quality and variance. Finally, our algorithms are efficient
and applicable to large uncertain graphs.

\section{Conclusion} \label{sec:conc}
Sparsification has often been used to reduce the size of deterministic graphs and facilitate efficient query processing. However, it has not been applied previously to uncertain graphs, despite the fact that they incur significantly higher cost for common query and mining tasks. This paper introduces novel sparsification techniques that, given an uncertain graph $\mathcal{G}=(V,E,p)$ and a parameter $\alpha \in (0,1)$, they return a subgraph $\mathcal{G}' = (V,E',p')$, such that ${E' : E' \subset E, |E'| = \alpha|E|}$. $\mathcal{G}'$ preserves the structural properties of $\mathcal{G}$, has less entropy than $\mathcal{G}$, and can approximate the result of various queries on $\mathcal{G}$.

The proposed methods, \GDB\ (\textit{Gradient Descent Backbone}) 
and \EMD\ (\textit{Expectation Maximization Degree}), involve a 
two-step framework. First, a backbone deterministic graph $G_b$ 
with $\alpha |E|$ edges is generated. In order to obtain 
$\mathcal{G}'$, \GDB\ assigns probabilities to the edges of $G_b$ 
aiming at preserving the expected vertex degrees or cut sizes, 
while reducing the entropy. In addition to assigning probabilities, \EMD\ also changes the structure of $G_b$ by 
adding or removing edges. 
An extensive experimental evaluation with real and synthetic uncertain graphs confirms that \GDB\ and \EMD\ consistently outperform benchmarks adapted from the deterministic graph literature, on several graph queries and metrics.         

\section*{Acknowledgements}
This work was supported by GRF grant 16201615 from Hong Kong RGC.

\balance

\bibliographystyle{abbrv}

\newpage
\appendix
\begin{algorithm}[!ht]
  \caption{Nagamochi Ibaraki (\textsf{NI})}
  \label{algo:NI}
  \begin{algorithmic}[1]
    \Require graph $G_w=(V,E,w)$, approximation parameter $\epsilon$
    \Ensure sparse graph $G'_w=(V,E',w')$
%
	\State $E'\leftarrow \emptyset$; $E_c\leftarrow E$; $F_0=\emptyset$
	\State $r = 0$
	\While {$E_c\neq \emptyset$}
		\State $r\leftarrow r+1$
		\State \begin{varwidth}[t]{\linewidth}
			compute a spanning forest $F_r$ of $E_c$ such that \par
			$(F_r\setminus F_{r-1})\cap E_c=\emptyset$ 
			\end{varwidth}
		\ForAll {edge $e\in F_r$ }
			\State $w_e \leftarrow w_e -1$
			\If {$w_e=0$}
				\State \begin{varwidth}[t]{\linewidth} 
				sample $e$ with probability \par
				$\ell_e = \min{\{\log|V|/(\epsilon^2\cdot r),1\}}$ \label{line:NI_add}
				\end{varwidth}
				\If {$e$ is sampled}
					\State $E'\leftarrow E' \cup \{e\}$ with $w'_e\leftarrow w_e/\ell_e$
				\Else 
					\State discard $e$
				\EndIf
				\State $E_c\leftarrow E_c \setminus\{e\}$
			\EndIf
		\EndFor
	\EndWhile
  \end{algorithmic}
\end{algorithm}

This section provides details of the benchmark methods \NI\ \cite{nagamochi1992linear} and \SS\ \cite{baswana2007simple}.  
Algorithm \ref{algo:NI} contains the core iterative process of \NI. The method requires as input an approximation parameter $\epsilon$, which is initially tuned depending on our parameter  $\alpha$; $\epsilon = \sqrt{|V|\log^2|V|/\alpha|E|}$. The result set $E'$ is initially empty and the set of available edges $E_c\gets E$. At each iteration $r$, a spanning forest $F_r$ is computed on the set $E_c$ with the requirement that if an edge $e$ appears in $F_{r-1}$, then it must also appear in $F_r$ (contiguous spanning forests). Each time an edge $e$ is covered by a spanning forest, $w_e$ is reduced by one. When $w_e$ becomes 0, $e$ is sampled with probability $\ell_e = \min{\{\log|V|/(\epsilon^2\cdot r),1\}}$. If $e$ is selected, then line \ref{line:NI_add} adds it to the result set. Otherwise, $e$ is discarded. The iterative process stops when all edges of $E$ have been examined. Intuitively, the sampling probability $\ell_e$ approximates the connectivity of edge $e$; if $e$ belongs to a sparse subgraph, it is covered by  spanning trees of early iterations, therefore it is sampled with high probability. On the other hand, an edge in a dense component is covered after several iterations $r$, thus it's sampling probability is significantly smaller.

Algorithm \ref{algo:SP} presents the main process of \SS\ \cite{baswana2007simple}, which computes a $(2t-1)$-$spanner$ of $O(t\cdot m^{1+1/t})$ expected size. 
The algorithm performs $t-1$ iterations, incrementally forming clusters of vertices with the minimum edge that connects them. $C_i$ maintains the set of vertex clusters for iteration $i$. Initially, $C_i$ contains $|V|$ sets of individual vertices (line \ref{algo:SPinit}) and the spanner $E'$ is empty. At each iteration, a set of clusters, $R_i$, is selected with probability $n^{-1/t}$ for each cluster. For each vertex $u \notin R_i$ that is not yet connected to the spanner, \SS\ examines its neighbours. If none of them is in $R_i$,  then for each adjacent cluster $c \in C_{i-1}$, \SS\ adds the least weight edge to $E'$, and updates the clusters (line \ref{algo:SPadd}). If $u$ has an adjacent node in $R_i$, the least weight edge $e=(u,v \in R_i)$ is added to the spanner (line \ref{algo:SPadd1}). Then, \SS\ iterates over all edges of $u$, adding for each adjacent cluster $c$, the minimum weight edge $e'$, if  $w(e') < w(e)$ (line \ref{algo:SPadd2}). To ensure that the resulting spanner is connected, the algorithm joins all remaining clusters using the connecting edge with the minimum weight (lines \ref{algo:SPjoin1} - \ref{algo:SPjoin2}).

\begin{algorithm}[!b]
  \caption{Spanner Sparsification (\textsf{SS})}
  \label{algo:SP}
  \begin{algorithmic}[1]
    \Require uncertain graph $G=(V,E,w)$, $t$
    \Ensure sparse spanner $G'=(V,E',w)$

	\State $E' \gets \emptyset$ spanner edges
	\State $V_S \gets \emptyset$ spanner vertices
	\State $C_0= \{\{u\}| u \in V\}$ clusters \label{algo:SPinit}
	\For {i=1 to t-1}
		\State $R_i\leftarrow$ sample $C_{i-1}$ with probability $n^{-1/t}$
		\State $C_{i}\leftarrow R_i$
		\ForAll {$u \in V \setminus V_S$ and $u \notin R_i$}
			\State $N_R \gets N(u) \cap R_i$ neighbours of $u$ in $R_i$
			\If{$N_R \neq \emptyset$}
				\State $e\gets$ minimum weight edge $u, v \in N_R$ \label{algo:SPadd1}
				\State $E' \gets E' \cup e$, $V_S \gets V_S \cup u$
				\State $E \gets E \setminus E(u, N_R)$ 
				\State merge clusters $C_{i}(v), C_{i}(u)$ 
				\ForAll {cluster $c \in C_{i-1}$ and $c \notin R_i$}
					\State $e'\gets$ minimum weight edge 
					\If{$w(e') < w(e)$}
						\State $E' \gets E' \cup e'$ \label{algo:SPadd2}
						\State $E \gets E \setminus E(u,c)$  
						\State merge clusters $c, C_{i}(u)$
					\EndIf 
				\EndFor
			\Else
				\ForAll {cluster $c \in C_{i-1}$}
					\State $e\gets$ minimum weight edge $u, N(u) \cap c$ \label{algo:SPadd}
					\State $E' \gets E' \cup e$, $V_S \gets V_S \cup u$
					\State $E \gets E \setminus E(u,c)$  
					\State merge clusters $c, C_{i}(u)$
				\EndFor
			\EndIf
		\EndFor
	\EndFor
	\ForAll {$c \in C_{t-1}$} \label{algo:SPjoin1}
		\State $e\gets$ minimum weight edge $c, v \in N(c)$
		\State $E' \gets E' \cup e$
	\EndFor \label{algo:SPjoin2}

  \end{algorithmic}
\end{algorithm}

Recall from Section 3.3 that, in both methods, the expected number of edges is given in $O$-notation, thus the sparsified graph is not guaranteed to reach $\alpha|E|$ edges. To ensure this, we run Algorithms \ref{algo:NI} and \ref{algo:SP} with modified parameters to approximate the smallest $\epsilon$ and $t$ that contains $E'<\alpha|E|$. 
The remaining $\alpha|E|-|E'|$ edges are sampled using the original probabilities.

\par

\end{document}